\pdfoutput=1

\documentclass[10pt,a4paper]{article}

\usepackage{fullpage}

\usepackage{amsmath,amssymb}

\usepackage{changepage}

\usepackage[utf8x]{inputenc}

\usepackage{textcomp,marvosym}

\usepackage{cite}

\usepackage{nameref,hyperref}

\usepackage{microtype}
\DisableLigatures[f]{encoding = *, family = * }

\usepackage[table]{xcolor}

\usepackage{array}

\usepackage{graphicx}

\newcolumntype{+}{!{\vrule width 2pt}}

\newlength\savedwidth

\newcommand{\beginsupplement}{%
        \setcounter{table}{0}
        \renewcommand{\thetable}{S\arabic{table}}%
        \setcounter{figure}{0}
        \renewcommand{\thefigure}{S\arabic{figure}}%
     }

\usepackage[aboveskip=1pt,labelfont=bf,labelsep=period,justification=raggedright,singlelinecheck=off]{caption}

\makeatletter
\renewcommand{\@biblabel}[1]{\quad#1.}
\makeatother

\date{}

\begin{document}
\vspace*{0.2in}

\begin{flushleft}
{\Large
\textbf\newline{Identifying stochastic oscillations in single-cell live imaging time series using Gaussian processes} 
}
\newline
\\
Nick E. Phillips\textsuperscript{1\textcurrency},
Cerys Manning\textsuperscript{1},
Nancy Papalopulu\textsuperscript{1*},
Magnus Rattray\textsuperscript{1*}
\\
\bigskip
\textbf{1} Faculty of Biology, Medicine and Health, University of Manchester, Manchester, United Kingdom
\\
\bigskip

%
%


\textcurrency Current Address: The Institute of Bioengineering, School of Life Sciences, Ecole Polytechnique F\'ed\'erale de Lausanne, Lausanne, Switzerland 

* Nancy.Papalopulu@manchester.ac.uk, Magnus.Rattray@manchester.ac.uk

\end{flushleft}
\section*{Abstract}
Multiple biological processes are driven by oscillatory gene expression at different time scales. Pulsatile dynamics are thought to be widespread, and single-cell live imaging of gene expression has lead to a surge of dynamic, possibly oscillatory, data for different gene networks. However, the regulation of gene expression at the level of an individual cell involves reactions between finite numbers of molecules, and this can result in inherent randomness in expression dynamics, which blurs the boundaries between aperiodic fluctuations and noisy oscillators. This underlies a new challenge to the experimentalist because neither intuition nor pre-existing methods work well for identifying oscillatory activity in noisy biological time series. Thus, there is an acute need for an objective statistical method for classifying whether an experimentally derived noisy time series is periodic. Here, we present a new data analysis method that combines mechanistic stochastic modelling with the powerful methods of non-parametric regression with Gaussian processes. Our method can distinguish oscillatory gene expression from random fluctuations of non-oscillatory expression in single-cell time series, despite peak-to-peak variability in period and amplitude of single-cell oscillations. We show that our method outperforms the Lomb-Scargle periodogram in successfully classifying cells as oscillatory or non-oscillatory in data simulated from a simple genetic oscillator model and in experimental data. Analysis of bioluminescent live-cell imaging shows a significantly greater number of oscillatory cells when luciferase is driven by a {\it Hes1} promoter (10/19), which has previously been reported to oscillate, than the constitutive MoMuLV 5' LTR (MMLV) promoter (0/25). The method can be applied to data from any gene network to both quantify the proportion of oscillating cells within a population and to measure the period and quality of oscillations. It is publicly available as a MATLAB package.

\section*{Author summary}
Technological advances now allow us to observe gene expression in real-time at a single-cell level. In a wide variety of biological contexts this new data has revealed that gene expression is highly dynamic and possibly oscillatory. It is thought that periodic gene expression may be useful for keeping track of time and space, as well as transmitting information about signalling cues. Classifying a time series as periodic from single cell data is difficult because it is necessary to distinguish whether peaks and troughs are generated from an underlying oscillator or whether they are aperiodic fluctuations. To this end, we present a novel tool to classify live-cell data as oscillatory or non-oscillatory that accounts for inherent biological noise. We first demonstrate that the method outperforms a competing scheme in classifying computationally simulated single-cell data, and we subsequently analyse live-cell imaging time series. Our method is able to successfully detect oscillations in a known genetic oscillator, but it classifies data from a constitutively expressed gene as aperiodic. The method forms a basis for discovering new gene expression oscillators and quantifying how oscillatory activity alters in response to changes in cell fate and environmental or genetic perturbations.


\section*{Introduction}

Oscillatory dynamics are widespread in biology over a range of time scales, from the circannual migration and reproduction patterns of animals \cite{Gwinner2003} to cytosolic calcium oscillations on the order of seconds \cite{Dupont2011}. In between these extremes lie a plethora of oscillatory phenomena \cite{Goldbeter2012}, from circadian rhythms \cite{Nagoshi2004, Barkai1999, Zhang2014}, the cell cycle \cite{Ferrell2011, Tyson2008, Bieler2014, Feillet2014}, NF-KB oscillations in inflammation \cite{Nelson2004} and p53 oscillations in response to DNA damage \cite{Geva-Zatorsky2006}. Biological oscillations have attracted intense research focus due to the functional benefits conferred by oscillatory dynamics for cellular functions. The most intuitive advantage of oscillations is to function as clocks to measure time within a cycle. For example, the circadian clock is used to anticipate conditions and synchronise physiological responses to daily environmental changes \cite{Rensing2001}. Biological oscillators can be coupled to an output with much slower dynamics, such that every oscillation increases the output in a step-like manner \cite{Levine2014, Goodfellow2014}. They can also be used to measure distance, such as in the segmentation clock where oscillations in gene expression are used to control the formation of somites along the elongating vertebrate body axis, which are the precursors of the segmented vertebral column \cite{Oates2012}. 

In addition to measuring time and space, oscillations have attracted interest due to their unique properties of encoding information \cite{Sonnen2014}. Oscillations are defined by both amplitude and frequency, and manipulation of either can be used to transmit and process information \cite{Berridge1997, Micali2015}. For example, driving the Ras/ERK signalling pathway at different frequencies leads to downstream genes showing varied activation strengths, and hence transcription factor dynamics are decoded into multiple, distinct gene expression programs \cite{Toettcher2013}. The mechanistic basis of signal decoding is thought to be the activation timescale of downstream gene targets, which is regulated by the timescales of nucleosome remodelling \cite{Hansen2013, Hansen2015a}.

As new potential gene expression oscillators are discovered it is essential to have tools that can objectively classify a biological time series as rhythmic or arrhythmic. This is particularly challenging in data from individual cells because mRNA and protein production and degradation are controlled by random collisions between finite numbers of interacting molecules within the cell \cite{Phillips2016,McKane2005}, which creates noise known as intrinsic stochasticity. Other sources of stochasticity in oscillatory gene expression include variability in cellular parameters (extrinsic noise; \cite{Moore2015}) or interruption of cell-cell signalling \cite{Masamizu2006}. Live imaging shows that oscillations in single cells may display variable inter-peak distances and gradual shifts in phase over time (quasiperiodicity), such that the signal will not match a perfectly periodic sine wave \cite{Phillips2016,Imayoshi2013,Webb2016}. In addition, evidence suggests that gene transcription per se is a discontinuous process, where bursting episodes of transcription are interspersed with silent intervals \cite{Suter2011}, and this phenomenon can also produce fluctuating expression time series. Thus, intrinsic and extrinsic noise in single-cell expression dynamics makes it challenging to distinguish periodic (but stochastic) from aperiodic (but fluctuating) phenomena because they both produce noisy time series. Distinguising these two is important for three reasons: Firstly, biochemical oscillators are commonly generated by a combination of negative feedback, nonlinearity in reaction kinetics, time delay and appropriate matching of degradation time scales \cite{Novak2008,Woods2016}, while bursting transcription can be produced by switching between active and inactive promoter states \cite{Munsky2012}. Identifying oscillatory activity is therefore the first step to discovering the underlying mechanism generating a given fluctuating gene expression time series. Secondly, the cell may use oscillations for encoding information \cite{Sonnen2014} while it may try to counteract noise introduced by bursting transcription \cite{BaharHalpern2015a}. Thus, distinguishing the two phenomena is the first step for the experimentalist to understand how a cell may interpret the observed gene expression dynamics. Finally, genetic or molecular manipulations may cause oscillatory gene expression to cease and revert to aperiodic fluctuations. For example, {\it Hes1} stops oscillating when miR-9 is overexpressed \cite{Bonev2012} and other proneural genes stop oscillating when cells differentiate \cite{Imayoshi2013}. Here, again, it is important to be able to distinguish oscillations from aperiodic noise with some certainty in order to be confident that a change in the dynamics has occurred. We anticipate that the need for statistical methods to distinguish oscillatory from non-oscillatory gene expression will increase in the near future as bioinformatics methods are developed to identify oscillatory expression in single cell RNA-seq data \cite{Leng2015} and as genomic editing (e.g. CRISPR/cas9) methods allow the efficient generation of reporter fusion knock-ins.

Many methods for analysing biological time series are designed for estimating the period of known oscillators, such as encountered, for example, in circadian time series (reviewed in \cite{Zielinski2014}). A common and well-known method is the Fourier transform (and the power spectrum: the Fourier transform squared), which finds periodicity within a time series by matching (convolving) the signal with sine waves. One such example is the Fast Fourier Transform Nonlinear-Least-Squares algorithm (FFT-NNLS), which uses a Fourier transform to provide an initial guess of the period before fitting the signal as a sum of sinusoidal functions with a non-linear least squares procedure \cite{Plautz1997}. The Fast Fourier Transform can be inaccurate for period determination due to poor resolution in the frequency domain. Other analysis pipelines, such as spectrum re-sampling, use bootstrap methods within the power spectrum to refine the period estimate and obtain confidence intervals \cite{Costa2013}.

In addition to the Fourier transform, time series analysis using wavelets is also commonly used to estimate the amplitude and period of oscillations \cite{Webb2016,Moore2015}. Fourier based techniques assume that the time series is stationary, so the statistical properties such as mean, variance, autocorrelation, period etc. remain constant over time. Wavelet analysis does not assume stationarity and is therefore able to detect amplitude and period changes over time. However, the Fourier transform, the wavelet transformation and related period estimation techniques \cite{Zielinski2014} do not form a statistical test to classify whether a time series is periodic.

Algorithms for assessing statistical confidence that the time series is periodic (reviewed in \cite{Wu2014}) include methods such as JTK and RAIN \cite{Hughes2010, Thaben2014}. JTK and RAIN are non-parametric algorithms that organise the data into groups belonging to either the rising or falling part of an oscillation period. In JTK a statistical test compares the orderings of the data with those derived from a library of guessed waveforms (such as sine curve). An advantage of using only the rankings of the data points is that the method is highly resistant to outliers. RAIN tests both the rising and falling parts separately and is able to detect arbitrary, non-symmetric waveforms. However, both methods require prior knowledge of the expected period and assume that the peak-to-peak time of oscillations is constant. While they may be useful for the circadian clock, where the period is known to be approximately 24 hours, they do not form a general test that can be applied to any biological system. The Lomb-Scargle Periodogram (LSP) is another method that can be used in objectively classifying data as oscillatory or not \cite{Lomb1976, Scargle1982, Glynn2006}. The LSP is a least-squares fit of the data to sinusoidal curves and unlike the Fourier transform it can handle non-evenly spaced and missing data. The null distribution of the LSP from aperiodic white noise is known, and hence for every frequency the LSP provides the probability that the contribution could have just been produced just by white noise. It does not require a previously known period and performs well when benchmarked against other popular competing schemes \cite{Zhao2008}.

A limitation of the LSP in applications to single-cell time series is that the assumed model of oscillatory and non-oscillatory dynamics may be unrealistic. The LSP and similar methods are typically benchmarked by generating a sine wave or other perfectly periodic waveform and adding white noise to simulate measurement error \cite{Zhao2008, Zielinski2014, Wu2014}. The competing methods are then tested by comparing their performance in detecting the periodicity in the simulated signals. However, the choice of waveforms is ad hoc, and there is no underlying mathematical model of the gene expression dynamics producing the oscillatory time series. Additionally, existing methods commonly assume that non-oscillatory expression is described by white noise and use this to form a null hypothesis for statistical testing \cite{Glynn2006}. The key assumption of white noise is that consecutive time series measurements are random and completely uncorrelated. However, all components of gene expression have a finite time scale of degradation such that if, for example, a protein is in high abundance at one time point, it is likely to be high at the next. This is poorly modelled by white noise, where by definition consecutive time points have no correlation. In summary, current statistical methods for detecting periodicity are not suitable for data emerging from recent single-cell imaging studies that are generated by stochastic networks of gene expression.

Another approach that is complementary to data driven approaches has been to employ ``bottom-up'' dynamical modelling, which takes a hypothesis of interactions within a gene regulatory network and predicts how the products of gene expression (mRNA, protein etc.) will evolve in time. Stochastic dynamical modelling accounts for the natural randomness inherent to single cells due to low copy number of interacting molecular components \cite{Schwanhausser2011}. Even though stochastic models are also subject to assumptions they none-the-less form a mechanistic expectation of dynamics at a single-cell level. There are a range of techniques to analyse a given time series that are based on stochastic modelling (reviewed in \cite{Bronstein2015}), from inferring parameters of a model of a transcription network (e.g. transcription rate) \cite{Heron2007} to inferring a sequence of promoter states and potential refractory periods in transcriptional activity \cite{Hey2015, Zechner2014}. Although these methods model the time series at a more mechanistic level, they once again do not output a confidence estimate as to whether the time series is periodic. 

Here we describe a new method that can be used as a statistical test to determine whether a given time series is periodic or not. To develop this new method we used a mechanistic (but general) model of intrinsically stochastic gene expression to define a forward model of what we expect oscillatory and non-oscillatory dynamics to look like, using Gaussian processes. Gaussian processes then allow us to perform data analysis, and we develop a statistical method to determine the confidence that a given single-cell time series is periodic based on our expectation of oscillatory and non-oscillatory dynamics. The new method has two critical advantages over previous methods for the analysis of single cell data: firstly, it can deal with a ``drifting phase'' that is often present in single cell gene regulation data, such that the exact peak-to-peak time of oscillations varies and the time series is therefore poorly described as a perfectly periodic signal. This peak-to-peak variation is naturally embedded in the newly presented analysis pipeline through use of a quasi-periodic covariance function. Secondly, our non-oscillatory model accepts that a time series can be randomly fluctuating and that consecutive time points can be correlated due to the finite degradation timescales of mRNA and protein. This is more likely to be a realistic model of non-oscillatory gene expression than white noise.

We demonstrate that our approach is more effective than the LSP at classifying cells as oscillatory or non-oscillatory in synthetic data generated from a stochastic model of gene expression. We then apply our method to experimental data, and show that significantly more cells are classified as oscillating in time series from the {\it Hes1} genetic oscillator than an Moloney murine leukaemia virus MoMuLV 5' LTR (MMLV) control, which is representative of fluctuating, non-oscillatory gene expression. Our analysis pipeline calculates the number of oscillating cells within a population and also parameters quantifying dynamic behaviour, such as period of oscillations. The coherence of oscillations can also be quantified through the quality Q-factor \cite{DEysmond2013,Webb2016}. It can be applied to single-cell live imaging gene expression data using reporters such as fluorescence or bioluminescence and can be used to both make comparisons between genes and to characterise changes in dynamics caused by perturbations and genetic mutations.

\section*{Methods}

The principles of the new method, which is described in detail below, are as follows: Firstly, we define a general model of gene regulation in terms of the underlying reactions of synthesis and degradation of molecular species in a single cell, whereby the copy number of the observed species changes with time. Then we apply the linear noise approximation (LNA) to the system to approximate the dynamics as a Gaussian process. In a Gaussian process model, measurements at different times follow a multivariate normal (Gaussian) distribution, which is defined only by a mean and a covariance function. The covariance function determines how correlated a pair of measurements is as a function of their separation in time. The LNA provides a theoretical model for two cases, depending on the covariance function: 1) intrinsically noisy oscillations, 2) random aperiodic fluctuations, which form the null model of non-oscillatory behaviour. These define a mechanistic, albeit abstract, model-based representation of single-cell gene expression dynamics. 

Stochastic modelling and the LNA is a forward modelling approach in that we can generate time series data from a given model. Our next step is to perform this process in reverse and infer the most likely model from a given time series, which can be derived from synthetic or experimental data. As the LNA describes a Gaussian process, the tools of non-parametric regression with Gaussian processes can be used to compute the probability of the data under a process with an oscillatory or non-oscillatory covariance function \cite{Rasmussen2006}. We can then provide a confidence estimate that the time series is periodic using the likelihood ratio between the oscillatory and non-oscillatory models. Finally, we use synthetic non-oscillatory data (defined by the appropriate covariance function) to choose a classification threshold for the likelihood ratio statistic in order to control the false discovery rate (FDR) at a specified significance level. MATLAB code to implement the method and reproduce all of the figures is available at https://github.com/ManchesterBioinference/GPosc. 

\subsection*{Defining a general model of gene expression dynamics}

We firstly define a general model of gene expression and then show how its dynamics can be approximated as a Gaussian process. The biochemical reactions controlling gene expression are the result of probabilistic encounters between discrete numbers of molecules. The exact order and timing of reactions are random, and this is mathematically described with the chemical master equation (CME). The CME formulates the stochastic kinetics in terms of underlying reactions, and captures the probability of having a certain number of molecules as a function of time. The CME is formulated from the underlying microscopic reactions of the system, where \emph{R} chemical reactions change the molecule number of \emph{N} chemical species

\begin{align}
s_{1j}X_{1} + ... + s_{Nj}X_{N} \overset{f(X)_{j}}\longrightarrow r_{1j}X_{1} + ... + r_{Nj}X_{N}
\end{align}
where \emph{j} is the index of the reaction number, $X_i$ is the chemical species $i$, $s_{ij}$ and $r_{ij}$ are the stoichiometric coefficients and $f(X)_{j}$ is the rate of reaction $j$, which depends on both the network of interactions and the kinetics of the reaction.
The probabilistic evolution of the system is described with the CME, and approximate solutions can be found using van Kampen's system-size expansion \cite{VanKampen1997,Grima2010}. The system-size expansion decomposes the time evolution of each chemical species into a deterministic and stochastic contribution, where the relative amplitude of stochastic fluctuations decrease with the inverse of the square root of the system size (i.e. relative fluctuations decrease with higher molecule number)
\begin{equation}
\label{ansatz}
\frac{n_i(t)}{\Omega} = x_i(t) + \Omega^{-1/2}\epsilon_i(t) \ ,
\end{equation}
where $x(t)$ is the deterministic solution, $\Omega$ is the system size and $\epsilon(t)$ describes the stochastic fluctuations around the deterministic solution. While $n_i$ represents the discrete copy number of reactants, $\epsilon_i$ can be modelled as a continuous random variable for large $n_i$ and $\Omega$. The ansatz (Eq~(\ref{ansatz})) can be used to derive the LNA (see Elf and Ehrenberg \cite{Elf2003} for details). When the deterministic equations reach a fixed-point, the LNA recovers a system of linear Langevin equations for the dynamics of the stochastic fluctuations
\begin{equation}
\frac{d}{dt} \boldsymbol{\epsilon}(t) = \mathbf{J}(\mathbf{x}^*)\boldsymbol{\epsilon}(t) + \sqrt \mathbf{D}(\mathbf{x}^*)\boldsymbol{\zeta}(t) \ ,
\label{langevin}
\end{equation}
\begin{equation}
\begin{split}
J_{iw} = \sum \limits_{j=1}^R S_{ij} \frac{\partial f_j(\mathbf{x}^*)}{\partial x_w} \ ,
\end{split}
\end{equation}
\begin{equation}
\begin{split}
D_{ik} = \sum \limits_{j=1}^R S_{ij} S_{kj} f_j(\mathbf{x}^*) \ ,
\end{split}
\end{equation}
where $J$ represents the Jacobian of the deterministic dynamics evaluated at steady state, $D$ represents the diffusion matrix, and $\zeta(t)$ is Gaussian white noise with $\langle \zeta_i (t)\zeta_j(t') \rangle = \delta_{ij} \delta(t-t')$. The stoichiometric matrix, $S$, describes how molecule numbers of each species are changed by reactions, and it is given by $S_{ij} = r_{ij} - s_{ij}$. Eq~(\ref{langevin}) describes the dynamics of noisy fluctuations around a deterministic steady-state. It defines a Gaussian process, where the fluctuations $\boldsymbol{\epsilon}(t)$ are distributed as a multivariate normal distribution. A multivariate normal distribution is fully described by its mean and covariance. As the deterministic equations are at a steady state, the mean of the process is zero. The covariance function $\mathbf{K}(\tau)$ can be solved analytically, as the Langevin equations derived through the LNA are linear (a multivariate Ornstein Uhlenbeck (OU) process) \cite{Gardiner2009},

\begin{equation}
\mathbf{K}(\tau) = \boldsymbol{\sigma} \exp^{\mathbf{J}\tau}.
\end{equation}
where $\tau = |t - t'|$ is the time difference between two points. The quantity $\sigma$ satisfies
\begin{equation}
\mathbf{J}\boldsymbol{\sigma} + \boldsymbol{\sigma} \mathbf{J}^{T} + \mathbf{D} = 0 \ .
\end{equation}
\subsection*{Modelling single cell oscillations and aperiodic random fluctuations}

Eq~(\ref{langevin}) defines a general representation of single-cell gene expression dynamics approximated as a Gaussian process, and from this general representation we now define models of single-cell oscillations and aperiodic random fluctuations. The Jacobian contains information on all of the parameters of the model and is dependent on the network interactions, which may not be known. To simplify the covariance function, the Jacobian can be diagonalised by the transformation $UJU^{-1} = diag(\lambda_1,\lambda_2,...,\lambda_N)$, and the covariance matrix is transformed to $\widetilde{K}(\tau) = U K(\tau) U^\dag$. The covariance function for eigenvector $i$ is then

\begin{equation}
\widetilde{K}_i(\tau) = \widetilde{\sigma}_i \exp^{\lambda_i \tau}.
\end{equation}

The eigenvalue $\lambda_i$ can either be real or complex, and we therefore define two covariance functions, labelled ``OU" and ``OUosc"

\begin{align}
{K_{OU}}(\tau) &= \tilde{\sigma}_{OU} \exp({-\alpha \tau}) \ , \quad & \textnormal{OU} \\
{K_{OUosc}}(\tau) &= \tilde{\sigma}_{OUosc} \exp({-\alpha \tau})\cos({\beta \tau}) \ . \quad & \textnormal{OUosc}
\end{align}

The OU covariance function represents a process with aperiodic stochastic fluctuations and no peak in the power spectrum \cite{Gardiner2009}. The OU covariance function has two parameters: $\sigma_{OU}$, quantifying the variance of the generated signal, and $\alpha$, which describes how rapidly the time series fluctuates. Illustrative examples are provided below.
The OUosc covariance function defines a quasi-periodic oscillatory process which captures the peak-to-peak variability inherent to stochastic biochemical oscillators. For simplicity we have only taken the real component of the eigenvector covariance function, and hence there is only a term proportional to $\exp({-\alpha \tau})\cos({\beta \tau})$, although in principle an even more general model could include an additional contribution proportional to $\exp({-\alpha \tau})\sin({\beta \tau})$. Note that the OUosc covariance function has been similarly derived for a 2-dimensional underdamped oscillator modelled with linear langevin equations \cite{Westermark2009}, but was not used for likelihood-based inference in that work. Like the OU covariance function, OUosc similarly contains both $\sigma_{OUosc}$ and $\alpha$ parameters quantifying the variance and time scale of fluctuations, respectively, but it also contains a $\cos({\beta \tau})$ term. The covariance function displays damped oscillations at frequency $\beta$, and the system therefore undergoes stochastic oscillations at this period. Crucially, the OUosc covariance function is damped, and therefore peaks in the time series will only be locally correlated in time. For a given position of a peak in the time series, the timing of the next peak will be relatively well known, but subsequent peaks become increasingly difficult to predict, depending on the length scale of the dampening $\alpha$. The dampening of correlation over time leads to the phase drift and peak-to-peak variability within single-cell oscillators.

\begin{figure}[!h]
\begin{center}
\includegraphics{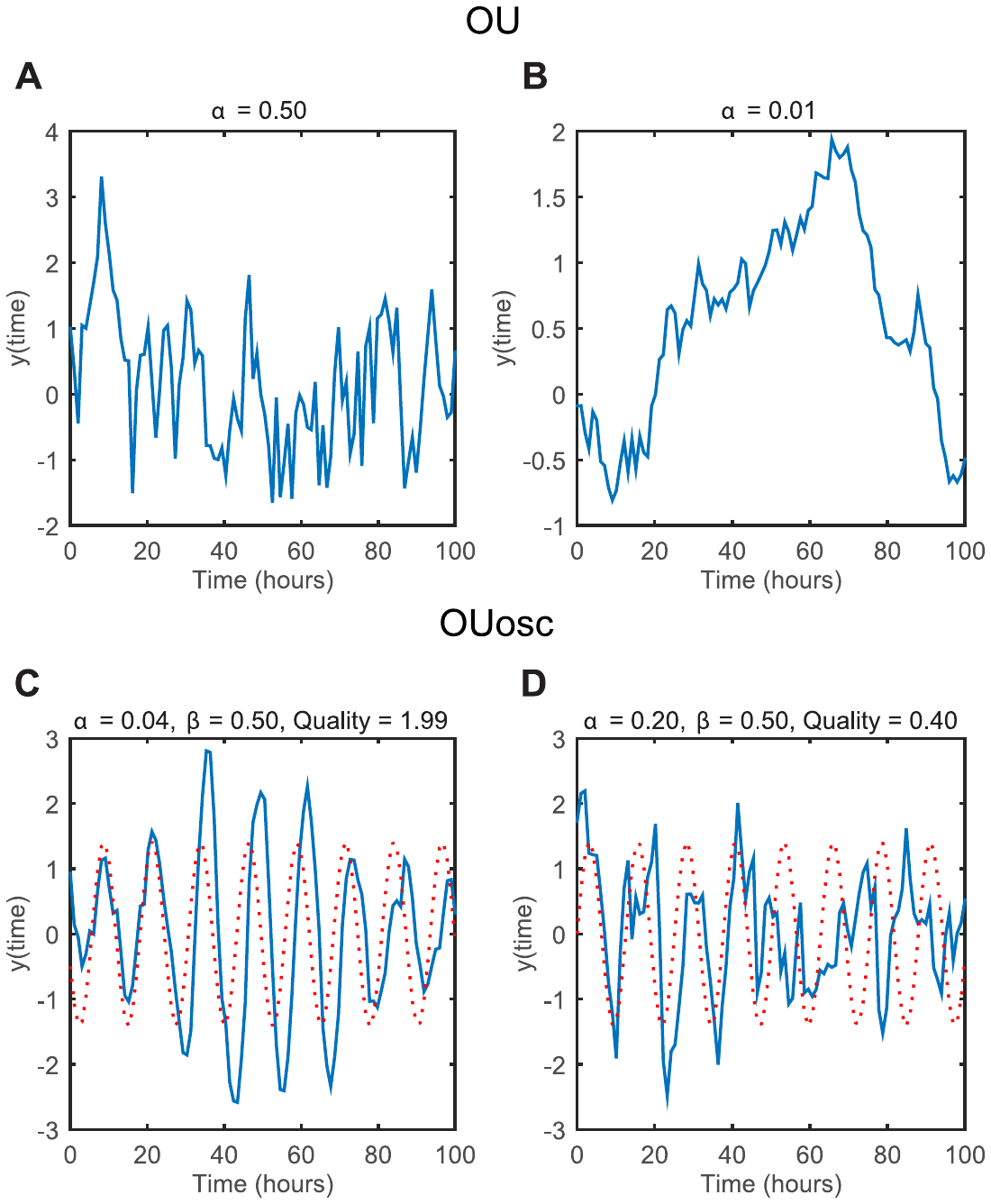}
\end{center}
\caption{{\bf Simulated time series examples from the OU and OUosc covariance functions in one dimension.} (A) Simulation of OU covariance function with $\alpha = 0.5$, $\sigma = 1$. (B) Simulation of OU covariance function with $\alpha = 0.01$, $\sigma = 1$. (C) Simulation of OUosc covariance function with $\alpha = 0.04$, $\beta = 0.5$, and $\sigma = 1$. (D) Simulation of OUosc covariance function with $\alpha = 0.2$, $\beta = 0.5$, and $\sigma = 1$. Dotted lines in (C) and (D) are sinusoids with frequency $\beta = 0.5$, where the first peak is fitted to the first peak of the data.}
\label{Examples}
\end{figure}

To visualise the time series generated by Gaussian processes with OU and OUosc covariance functions, Fig \ref{Examples} shows realisations with different parameter values. The amplitude (variance) of fluctuations is controlled by $\sigma$, which changes the scale bar, but does not change the dynamical properties; this is kept at 1 for all examples. For the non-oscillatory OU covariance function, the $\alpha$ parameter controls how long fluctuations are correlated in time. When $\alpha$ is high the covariance function is heavily damped and successive time points have low correlation (Fig \ref{Examples}A). As the correlation is low, future time points are highly unpredictable and hence the time series rapidly fluctuates. In contrast, when the damping parameter $\alpha$ is low, time points are correlated over longer time scales, as shown in Fig \ref{Examples}B. For the OUosc function, the $\beta$ parameter controls the frequency of oscillations while $\alpha$ similarly controls the rate of decay of the covariance function. Together, these parameters regulate the peak-to-peak variability and coherence of oscillations. The quality of oscillations can be quantified by the oscillation Q-factor, which quantifies the ratio of the frequency of oscillations to the timescale of damping \cite{DEysmond2013}

\begin{equation}
Q = \frac{\beta}{2\pi\alpha} \ .
\end{equation}

When the decay of the covariance is slow relative to the frequency of oscillations, oscillations are correlated for many cycles and there is low peak-to-peak variability, as shown in Fig \ref{Examples}C. The oscillations are of high quality (high Q-factor), and they reasonably match a sine wave at exactly the oscillatory frequency. However, when the quality is low (low Q-factor), the peak-to-peak variability becomes significant, and the signal is poorly described by a perfectly periodic sine wave (Fig \ref{Examples}D).
\subsection*{Gaussian process regression to classify cells as periodic or aperiodic}
Having derived stochastic models of single cell oscillations and aperiodic fluctuations, our goal is to assess whether a given time series is better described by a non-oscillatory OU or an oscillatory OUosc model. Firstly, we show that we can compute the probability of the time series data under both Gaussian process models. This allows us to define the log likelihood ratio (LLR) of the two models, which is used for model selection between the non-oscillatory OU and the oscillatory OUosc model and provides a level of confidence that a time series is periodic.

As the LNA describes a Gaussian process, the tools of non-parametric regression with Gaussian processes can be used to determine the likelihood function for both the OU and OUosc models \cite{Rasmussen2006}. Simulating a Gaussian process time series with either the OU or OUosc covariance function and measuring the value at times $\mathbf{t}$ will generate a random vector $f(\mathbf{t})$. Due to the randomness of the process, however, this represents just one sample from an infinite number of trajectories that could be created. Gaussian process regression is so powerful because it is analytically tractable to state the probability of the data being generated by a given model, even though there are infinite possible trajectories that the stochastic model could take. By integrating (marginalising) over all possible trajectories, the marginal likelihood of the data set $\mathbf{y}$ for a given model is

\begin{equation}
\log p(\mathbf{y}|\mathbf{t},\mathbf{\theta}) = -\frac{1}{2} \mathbf{y}^T K_y^{-1}\mathbf{y} - \frac{1}{2}\log |K_y| - \frac{n}{2} \log 2\pi,
\label{loglik}
\end{equation}

where $\theta$ are the hyperparameters of the models, which in the current context is $\alpha$, $\beta$ and $\sigma$. The covariance matrix $K$ has elements from the $K_{OU}$ or $K_{OUosc}$ covariance function evaluated at the time points where the data are collected. The first term describes how well the model fits the data, whereas the second term penalises model complexity of the covariance function. If the covariance function is rapidly damped (high $\alpha$), then it describes a process with extremely short timescales. As the process fluctuates rapidly, it will be able to fit the data well regardless of whether it is a true representation of the dynamics, and this second term corrects for this overfitting. The third term is a normalisation constant. If the data are subject to measurement error, each observation $y(t)$ can be related to the underlying function of the cellular behaviour $f({t})$ through a Gaussian noise model

\begin{align}
y(t) = f(t) + \mathcal{N} (0,\sigma^2_n) \ ,
\end{align}
where $\sigma^2_n$ represents the variance of the measurement noise. The noise can be combined with the covariance function

\begin{align}
K^*(t,t') = K(t,t') + \sigma^2_n \delta(t,t') \ .
\label{EqWithNoise}
\end{align}
The task is then to find the maximal likelihood for both the OU and OUosc covariance functions, which is found by varying the hyperparameters to find an optimum. The partial derivative of the marginal likelihood with respect to hyperparameters is

\begin{equation}
\frac{\partial}{\partial \theta_j}\log p(\mathbf{y}|\mathbf{t},\mathbf{\theta}) = \frac{1}{2} \mathbf{y}^T K_y^{-1} \frac{\partial K_y}{\partial \theta_j}K_y^{-1}\mathbf{y} - \frac{1}{2} tr (K_y^{-1} \frac{\partial K_y}{\partial \theta_j}) \ .
\end{equation}

The maximum marginal likelihood is found when the derivative of the marginal likelihood with respect to hyperparameters becomes zero. The marginal likelihood was maximised for both OU and OUosc models using the MATLAB GPML toolbox \cite{Rasmussen2010}, and the maximum likelihood of the two models can then be compared. The difference in the log-likelihood of the two models is known as the log-likelihood ratio, and the LLR between the OU and OUosc models then provides a statistic to determine whether the data is oscillating as opposed to aperiodic random fluctuations. 

\subsection*{Detrending to identify oscillations embedded in long term trends}

As well as oscillatory activity, gene expression may exhibit long term trends. Without accounting for trends in the data, oscillations can become harder to detect because the signal is dominated by an aperiodic long term trend. The OUosc model describes oscillations with variability in amplitude and period, but the mean of the covariance function is zero. Over long time scales it is clear that the average of the signal generated by the OUosc model remains constant (\ref{LongTimeSeriesExample2}), and consequently experimental data with long term trends may be poorly described by the OUosc alone and oscillations that co-exist with long term trends could be missed.

\begin{figure}[!h]
\begin{center}
\includegraphics[scale=1]{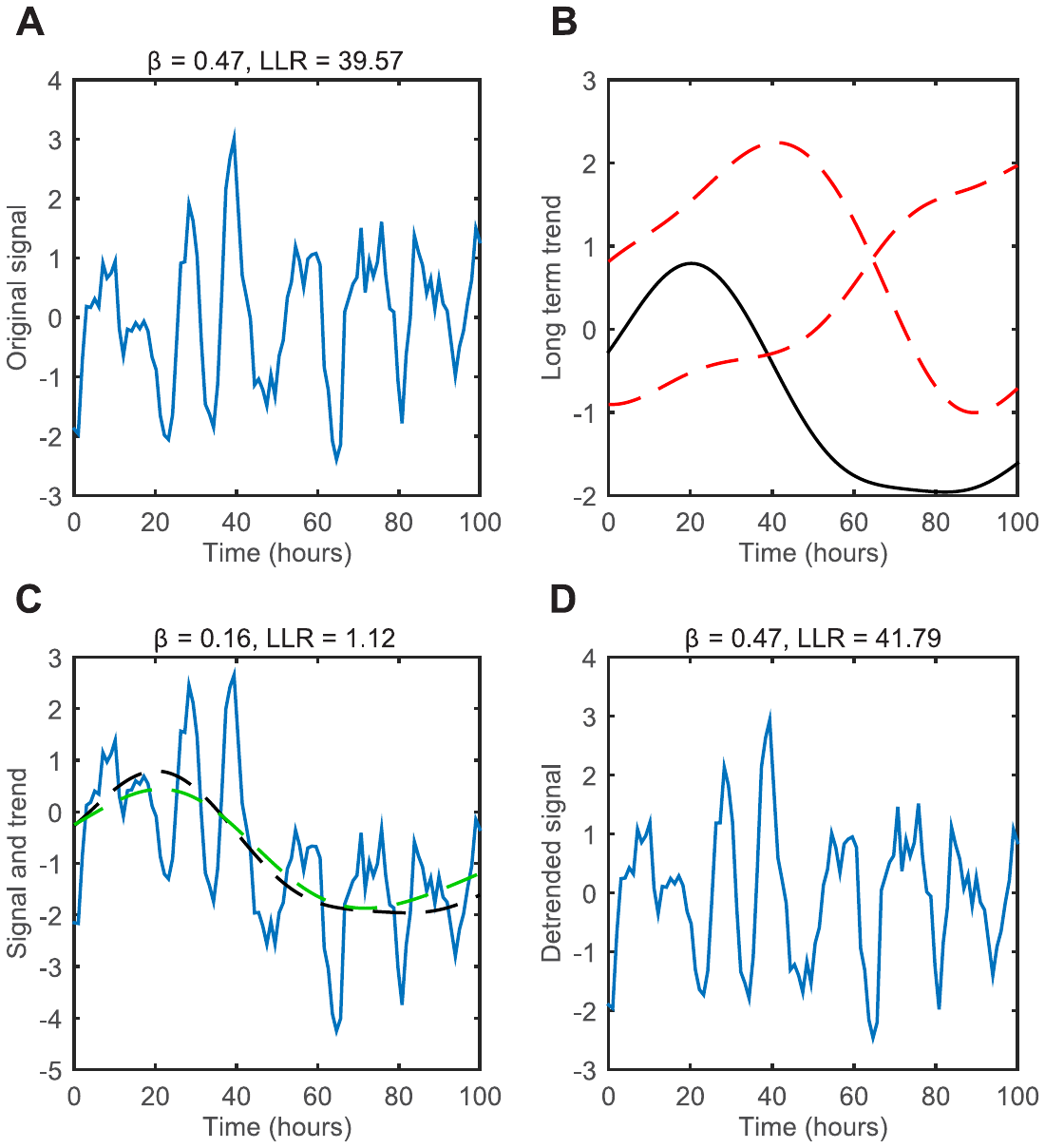}
\end{center}
\caption{{\bf An illustrative example of the detrending pipeline.} (A) Simulation sample from OUosc covariance function with $\alpha = 0.2$, $\beta = 0.5$, and $\sigma = 1$. (B) Simulation of 3 different samples of SE covariance function with $\alpha_{SE} = 0.01$ and $\sigma_{SE} = 1$. Solid black example is used then added to signal (C) Superposition of original signal (A) with long term trend ((B), shown as dotted black line). Dotted green line represents trend fitted from data. (D) Detrended final signal used for analysis, found by subtracting the fitted trend from (C). }
\label{Detrending}
\end{figure}

We therefore require a method to detrend the data and allow the detection of oscillations superposed on long term trends. Typically, detrending data involves fitting and removing a polynomial \cite{Westermark2009, Plautz1997}. Here we detrend using Gaussian processes, which has the advantage that we do not have to specify an exact functional form of the trend. We use the squared exponential (SE) covariance function, because it describes a general smooth process

\begin{align}
K_{SE}(t,t') = \sigma_{SE}\exp(-\alpha_{SE}\tau^2) \ , 
\end{align}
where $\tau = |t-t'|$ as before. The parameter $\alpha_{SE}$ can alternatively by expressed as a lengthscale of correlation, $l_{SE}$, with $l_{SE} = \sqrt{1/2\alpha_{SE}}$. Fig \ref{Detrending} illustrates how the addition of long term trends can interfere with the detection of oscillatory dynamics. Fig \ref{Detrending}A shows an example of a time series from an OUosc model, and the calculated LLR of 39.6 is strongly in favour of oscillations. Random functions drawn from the SE covariance function are then displayed in Fig \ref{Detrending}B, where the length scale of the trend is set by the parameter $\alpha_{SE}$. Crucially, when the long term trend is added to the original oscillatory signal, the calculated LLR markedly drops to 1.12, favouring the OU model (Fig \ref{Detrending}C). The essence of the detrending approach is to run this process in reverse, by first fitting an SE trend model to the data (Fig \ref{Detrending}C green line). By applying our detrending method and removing the fitted trend, the LLR of 41.8 once again indicates the presence of oscillations, as shown in Fig \ref{Detrending}D. When fitting the trend it is important to remove long term trends while preserving the oscillatory signal, and if the trend line is allowed to be too flexible it will fit both. To avoid this problem one can set an upper bound on $\alpha_{SE}$, such that there is a minimum length scale that the trend will remove. The importance of this upper bound is discussed in the results section. Note that both the trend and the OU/OUosc covariance functions can in principle be fitted simultaneously, but we found that optimisation of the hyper-parameters was often caught in poor local optima and therefore use a two-stage procedure.

\subsection*{Choosing an LLR threshold to classify cells and control false discoveries}
In order to classify a cell as oscillatory or non-oscillatory a threshold in the LLR must be reasonably chosen. The LLR between two models can be used in a log-likelihood ratio test to choose whether to reject the null (non-oscillatory) model in favour of the alternative model. When the models are nested, the test statistic is usually assumed to be asymptotically chi-squared distributed (using Wilks' theorem). However, Wilks' theorem assumes that the estimated parameters are in the interior of the more complicated model, and this does not apply here because the OU model is only equivalent to the OUosc model when the frequency is zero. We therefore use an empirical approach for choosing an LLR threshold to define a cut-off between oscillatory or non-oscillatory classification, that we now describe. 

An objective metric on which to base the LLR threshold is the false discovery rate (FDR). The FDR seeks to control the proportion of cells passing the oscillatory test that are actually non-oscillatory, but it differs from the false positive rate (FPR) often used in statistical testing. Given a particular LLR threshold, the FPR is the proportion of cells called significant (oscillatory) when they are actually null (non-oscillatory). The FPR is typically controlled by setting a significance level on the p-value, which is the probability of the observed test statistic under the null hypothesis (in our case the LLR). In contrast, the FDR quantifies the expected proportion of cells falsely called significant as a percentage of the total number of cells passing the threshold. So if, for example, out of 100 non-oscillatory cells the LLR threshold is chosen such that 5 are deemed significant, then the FPR is 5\%, but the FDR is 100\%, because all 5 of the cells called significant are actually false. The FDR therefore provides a better representation of the balance between the number of true positives and false positives in classifying cells.

Analogous to the p-value to control the FPR, the q-value is the minimum FDR attained at or above a given LLR score. For a particular LLR threshold, the associated q-value is the expected number of false positives as a proportion of all cells that exceed the threshold. To control the FDR and calculate q-values we follow the procedure proposed by Storey and Tibshirani \cite{Storey2003}. The FDR is quantified as
\begin{equation}
\begin{split}
\textnormal{FDR} &= E \left[\frac{\textnormal{number of false positives}}{\textnormal{total number of cells passing}}\right] \\
&\approx \frac{\textnormal{$\pi_0$ * FPR}}{\textnormal{total number of cells passing}} \ . 
\end{split}
\end{equation}
where $\pi_0$ is the estimated proportion of non-oscillating cells and FPR is the probability of non-oscillator passing. The method relies on first estimating the proportion of non-oscillating cells ($\pi_0$) by comparing the shape of p-values from the analysed data set with that expected from a population of non-oscillating cells. The original protocol uses the statistical properties of p-values, specifically that under the null hypothesis (i.e. a population of non-oscillating cells) the p-values are uniformly distributed between 0 and 1. As we calculate a LLR instead of a p-value, we cannot assume that the distribution of LLR scores is uniform (i.e. we do not know the expected distribution of LLR scores of non-oscillating cells). We therefore find this distribution through a bootstrap approach where we simulate many cells with the null aperiodic OU model and calculate the LLRs. In order to choose parameters to simulate cells we use the parameters of the OU model fitted to the data. Sampling from the null models to compute the test statistic is known as a Cox test \cite{Goldman1993, Cox1962}. The procedure of calculating q-values is as follows:

1) Calculate the LLR score for each cell in the data set. For each cell in the data set to be analysed (either experimental or computationally generated), fit an oscillatory (OUosc) and non-oscillatory (OU) Gaussian process model to the detrended data, and calculate the LLR difference between the fits of the two models. The LLR is normalised by dividing by the length of the data and multiplying by 100 (to make units more rounded). Let $LLR^{d}_{1} \leq LLR^{d}_{2} \leq ... \leq LLR^{d}_{m} $ represent the ordered $LLR$ values of the time series from the data.

2) Create a synthetic data set to approximate the LLR distribution expected from the null (OU) model. Generate 2000 synthetic cells by sampling equally from the parameters fitted by the trend and OU non-oscillatory model for each cell. If the data set consists of 100 cells, then the fitted trend and OU parameters of each cell would be used to create 20 synthetic cells, for example. Let $LLR^{s}_{1} \leq LLR^{s}_{2} \leq ... \leq LLR^{s}_{m’} $ represent the ordered $LLR$ values of the synthetic time series.

3) Estimate the proportion of non-oscillating cells by comparing the shape of the data set with that of non-oscillating cells generated in (2). For a range of $\lambda$, (e.g. $\lambda = 0.1, 0.15, ..., 1$ between minimum and maximum LLRs), estimate proportion of cells in the data set that are non-oscillatory
\begin{equation}
\pi_0 = \frac{ \# \{LLR^{d}_i < \lambda; i = 1,2,...,m\}/m}{\#\{LLR^{s}_i < \lambda; i = 1,2,...,m’\}/m’}
\end{equation}
where \# denotes the number of cells. $\pi_0$ must be estimated by tuning the parameter $\lambda$. Allowing $\widehat{f}$ to be natural cubic spline with 3 degrees of freedom of $\pi_0(\lambda)$ on $\lambda$, set the estimate of $\pi_0$ to be
\begin{equation}
\pi_0 = \widehat{f}(\textnormal{min}(\lambda))
\end{equation}

4) Calculate the q-value of each cell in the data set. 
\begin{equation}
\widehat{q}(LLR^D_m) = \min_{t \geq LLR^D_m} \frac{\widehat{\pi_0} \# \{LLR^{s}_j \geq t\}/m’}{\#\{LLR^{d}_j \geq t\}/m}
\end{equation}
For $i = m-1,m-2,...,1$ calculate
\begin{equation}
\widehat{q}(LLR^D_i) = \min_{t \geq LLR^D_i} \frac{\widehat{\pi_0} \# \{LLR^{s}_j \geq t\}/m’}{\#\{LLR^{d}_j\geq t\}/m}
\end{equation}

By controlling the q-value at a certain threshold ($q<\gamma$) we are then able to quantify the number of oscillating and non-oscillating cells within the population.

\subsection*{Generating synthetic oscillatory and non-oscillatory data for evaluating classification performance}

In order to assess the performance of the method to discriminate periodic and aperiodic signals, we generated synthetic data from a stochastic model consisting of reactions between discrete numbers of molecules in a network with a known oscillatory and non-oscillatory regime. Specifically, we used a model of the {\it Hes1} genetic oscillator \cite{Monk2003, Galla2009}, which consists of negative autoregulation with delay. The network topology is illustrated in Fig 3, and it is comprised of the following four reactions

\begin{align}
&{M} \overset{\mu_m}\longrightarrow \emptyset \\
&{P} \overset{\mu_p}\longrightarrow \emptyset \\
&{M} \overset{{\alpha_p}}\longrightarrow {M} + {P} \\
&\emptyset \overset{f(n_p)}\Longrightarrow {M} \ .
\end{align}
The first two reactions correspond to mRNA and protein degradation, respectively, where $\mu_m$ and $\mu_p$ are parameters describing their degradation rates. The third reaction describes the production of HES1 protein through translation, and is dependent on both the number of mRNA molecules and the translation rate parameter $\alpha_p$. The final term represents the production of {\it Hes1} mRNA through the process of transcription. The negative repression of transcription by HES1 protein is modelled with the Hill function $f(n_p) = \alpha_m/(1+(n_p/\Omega P_0)^h)$, where $P_0$ and $h$ are constants representing the strength of negative repression and $\Omega$ represents the system size. The double arrow denotes that this reaction contains the total delay within the system, such that when the reaction is triggered at $t$, an mRNA molecule is not produced until $t+\tau$. The parameters of the model control whether the system undergoes oscillations or aperiodic fluctuations \cite{Galla2009,Brett2013}. Data was simulated in an oscillatory and non-oscillatory parameter regime using the delayed version of Gillespie's stochastic simulation algorithm \cite{Anderson2007, Gillespie1977}. To compare with a pre-existing method, we also applied the LSP to the data using the ``plomb'' function within MATLAB.

\begin{figure}[!h]
\begin{center}
\includegraphics[scale=1]{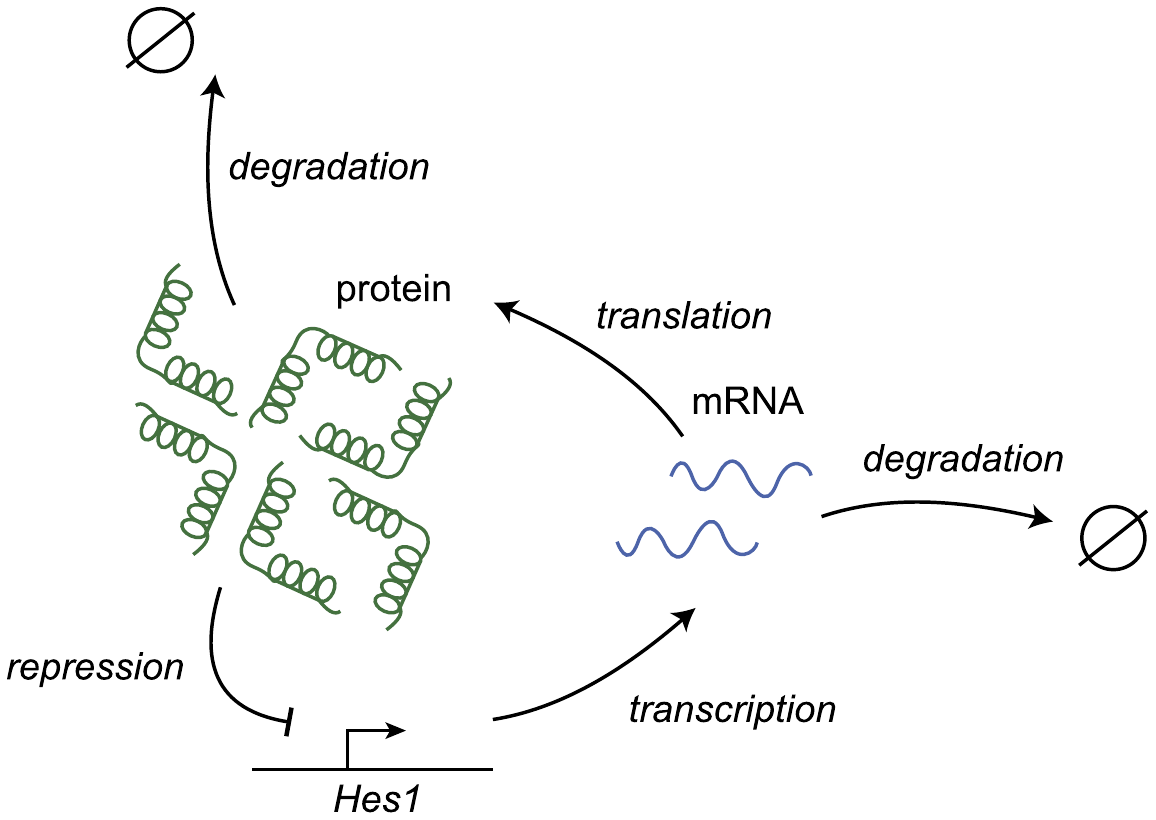}
\end{center}
\caption{{\bf The network topology of the {\it Hes1} transcription factor.}}
\label{Topology}
\end{figure}

\subsection*{Experimental procedures}
\subsubsection*{Cell-lines}
C17.2 cells were grown in DMEM (ThermoFisher Scientific, UK) with 10\% FBS (ThermoFisher Scientific, UK). The {\it Hes1} reporter construct contained 2.7kb of {\it Hes1} promoter upstream of destabilised ubiquitin-luciferase followed by the 540 bp {\it Hes1} 3’UTR \cite{Masamizu2006} within the pcDNA4 backbone plus Zeocin resistance gene. The control promoter construct comprised the pBabe puro retroviral backbone \cite{Morgenstern1990} containing the Moloney Murine Leukemia Virus Long Terminal Repeat (MMLV-LTR) upstream of destabilized ubiquitin-luciferase plus a Puromycin resistance gene.  Generation of stable {\it Hes1} reporter or control reporter cell-lines was performed by transfection of pcDNA4-{\it Hes1}::ubq-luciferase WT 3’UTR or pbabepuro::ubq- luciferase respectively into C17.2 cells using Lipofectamine 2000 (ThermoFisher Scientific, UK) and addition of 1 mg/ml Zeocin (ThermoFisher Scientific, UK) or puromycin (Sigma, UK) 5ug/ml after 48 hours. Cells were maintained in antibiotic selection for 2 weeks and individual resistant colonies picked to generate single-cell clones. C17.2 {\it Hes1}::ubq-luciferase clones were tested for luciferase expression and response to transient Notch1 Intra-cellular domain over-expression in a FLUOstar Omega plate reader (BMG LabTech, UK). A representative clone was used for subsequent imaging.
\subsubsection*{Bioluminescence imaging}
C17.2 reporter cells were plated on 35mm glass-based dishes (Greiner-Bio One, UK) and were allowed to adhere before serum withdrawal for 3 hours and subsequent imaging in the presence of 10\% serum and 1mM D-luciferin (Promega, UK). Plates were placed on an inverted Zeiss microscope stage and maintained at $37^{o}$C in 5\% CO2. Luminescent images were obtained using a 10x 0.3NA air objective and collected with a cooled charge-coupled device camera (Orca II ER, Hamamatsu Photonics). A 30 minute exposure and 2x2 binning was used.
\subsubsection*{Image analysis}
Bioluminescent movies were analysed in Imaris (Bitplane, UK). Images were first subject to a 3x3 median filter to remove bright spot artefacts from cosmic rays. Individual cells and background regions were tracked manually using the ``spots'' function and single cell bioluminescence values over time were extracted. Four background areas containing no cells were used to estimate and constrain the experimental noise parameter $\sigma_n$ in Eq~(\ref{EqWithNoise}).

\section*{Results}

In order to characterise the performance of the new method based on Gaussian processes we compare it with the LSP on synthetic data generated by a stochastic model of the {\it Hes1} genetic oscillator. We subsequently apply the method to live-cell reporter imaging from C17.2 neural progenitor cells to assess ability to discriminate dynamic gene expression generated by two different types of promoters: {\it Hes1} and MMLV.
\subsection*{Performance on synthetic data} 

Having defined a stochastic model to generate synthetic data, we now proceed to evaluate the performance of the new Gaussian process method versus the LSP \cite{Glynn2006} in classifying cells as oscillatory or non-oscillatory. The data is generated from a computational model of  the {\it Hes1} oscillator, which is a system of negative autoregulation with delay (see Fig 3). It is known a priori whether a time series generated by this model is periodic or aperiodic because it depends on the choice of parameters. Thus, we can use these synthetic data to quantify the statistical performance of the two methods.

\begin{figure}[!h]
\begin{center}
\includegraphics[scale=1]{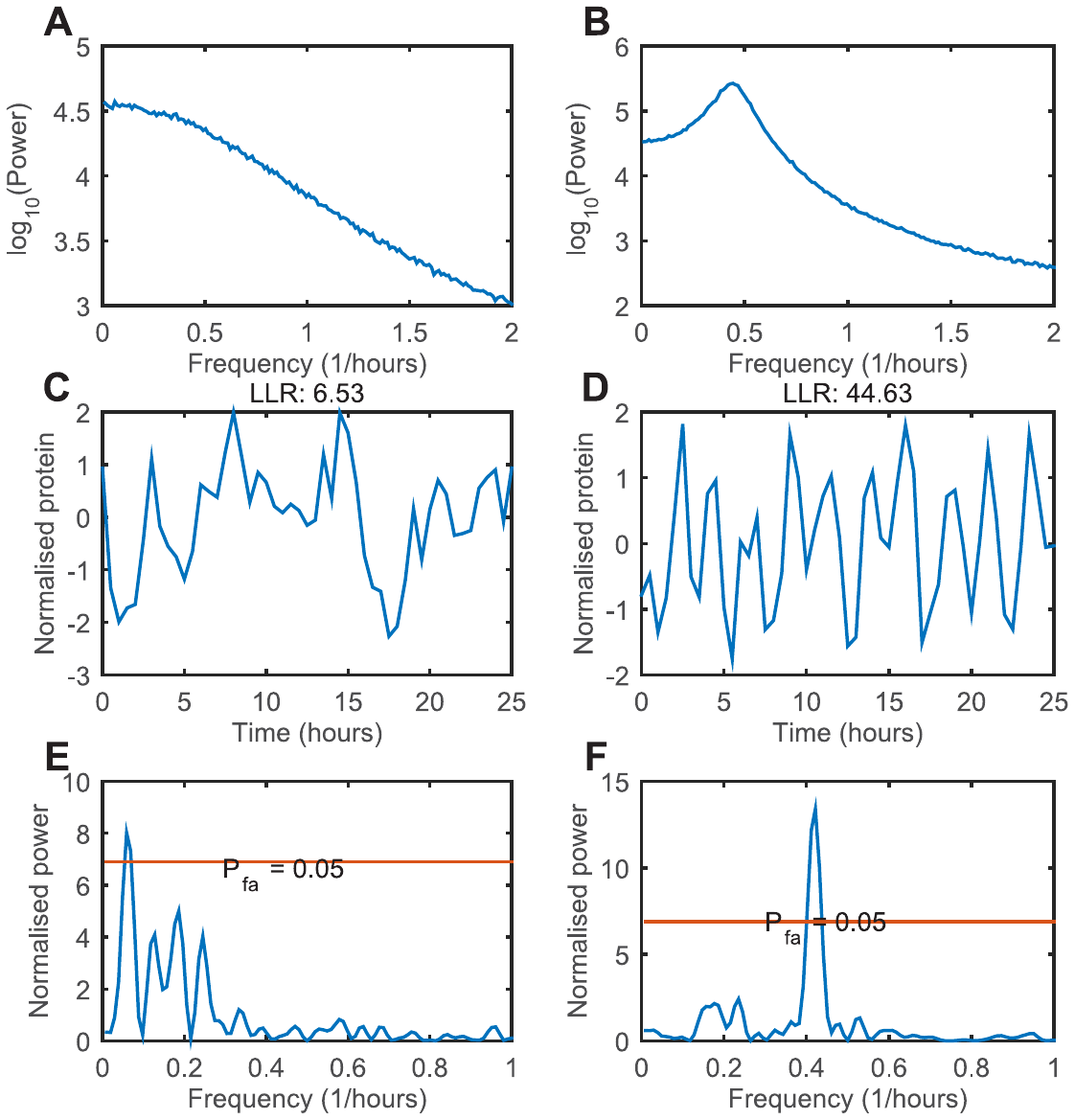}
\end{center}
\caption{{\bf Time series examples from the stochastic {\it Hes1} model}. (A), (B) Power spectra of protein concentrations calculated from 1000 independent Gillespie simulations. Measurements start at t = 5000 mins in order to allow for equilibration. Model parameters for (A) are $P_0$ = 300, $h$ = 1, $\tau$ = 0, $\alpha_m$ = $\alpha_p$ = 1, $\mu_m$ = $\mu_p$ = 0.07 and $\Omega$ = 20. Model parameters for (B) are $P_0$ = 100, $h$ = 3, $\tau$ = 18, $\alpha_m$ = $\alpha_p$ = 1, $\mu_m$ = $\mu_p$ = 0.03 and $\Omega$ = 20. (C), (D) Time series examples and associated LLR scores from (A) and (B), respectively. (E), (F) LSP of time series (C) and (D), respectively. }
\label{LS}
\end{figure}

In order to ascertain that the model is in an oscillatory or non-oscillatory regime, we calculate the average power spectra for cells simulated with the two different parameter sets (parameters given in Fig \ref{LS}). The power spectrum is the Fourier transform squared, and while is does not individually classify cells, it shows the average behaviour of a large population. For the non-oscillatory regime the average power spectrum of 1000 cells shows no peak (Fig \ref{LS}A), indicating that the cells undergo random fluctuations with no characteristic periodicity. The average power spectrum from the oscillatory regime shows a clear peak at a frequency of 0.5 hours$^{-1}$, corresponding to a period of 2 hours (Fig \ref{LS}B).

We simulated data from 1000 cells for the {\it Hes1} model in both the oscillatory and non-oscillatory parameter regimes, and the protein levels were measured every 30 minutes for 25 hours, which is approximately the same as available for experimental data \cite{Goodfellow2014}. Measurement noise with a variance of 10\% of the signal was then added to each time point. An example time series from the non-oscillatory regime is shown in Fig \ref{LS}C, and the calculated LLR for the cell was 6.5. Simulating 2000 synthetic cells with the OU model fitted to this data allows us to quantify the approximate p-value, which was calculated to be non-significant (p-value \textgreater 0.05). Fig \ref{LS}D shows an example from the oscillatory regime, with an associated LLR of 44.6. This value is greater than any produced by a synthetic population of 2000 OU non-oscillators and is hence classified as significant (upper bound p-value of $5 \times 10^{-4}$).

The LSP of the non-oscillating cell (Fig \ref{LS}C) is highly noisy and contains strong contributions at multiple frequencies (Fig \ref{LS}E). The signal exceeds the threshold for a significant p-value threshold of 0.05, and is therefore classified as oscillating. This particular cell would represent a false positive at this particular p-value threshold of 0.05. The LSP of the oscillating cell is shown in Fig \ref{LS}F, where the signal passes the 0.05 p-value significance threshold, and is classified as oscillating.

We use the Receiver Operating Characteristic (ROC) curve to systematically compare the performance of both methods \cite{Broadhurst2006}, which plots the true positive rate against the false positive rate of classifying oscillating cells for the synthetic data set. Both methods require a threshold to classify a cell as oscillatory (either LLR for Gaussian process or p-value for LSP), and varying the threshold controls the number of false positives to true positives. The ROC illustrates the performance of both methods as the threshold is varied, where a greater area under the curve indicates a better ratio of true positives to false positives.

\begin{figure}[!h]
\begin{center}
\includegraphics[scale=1]{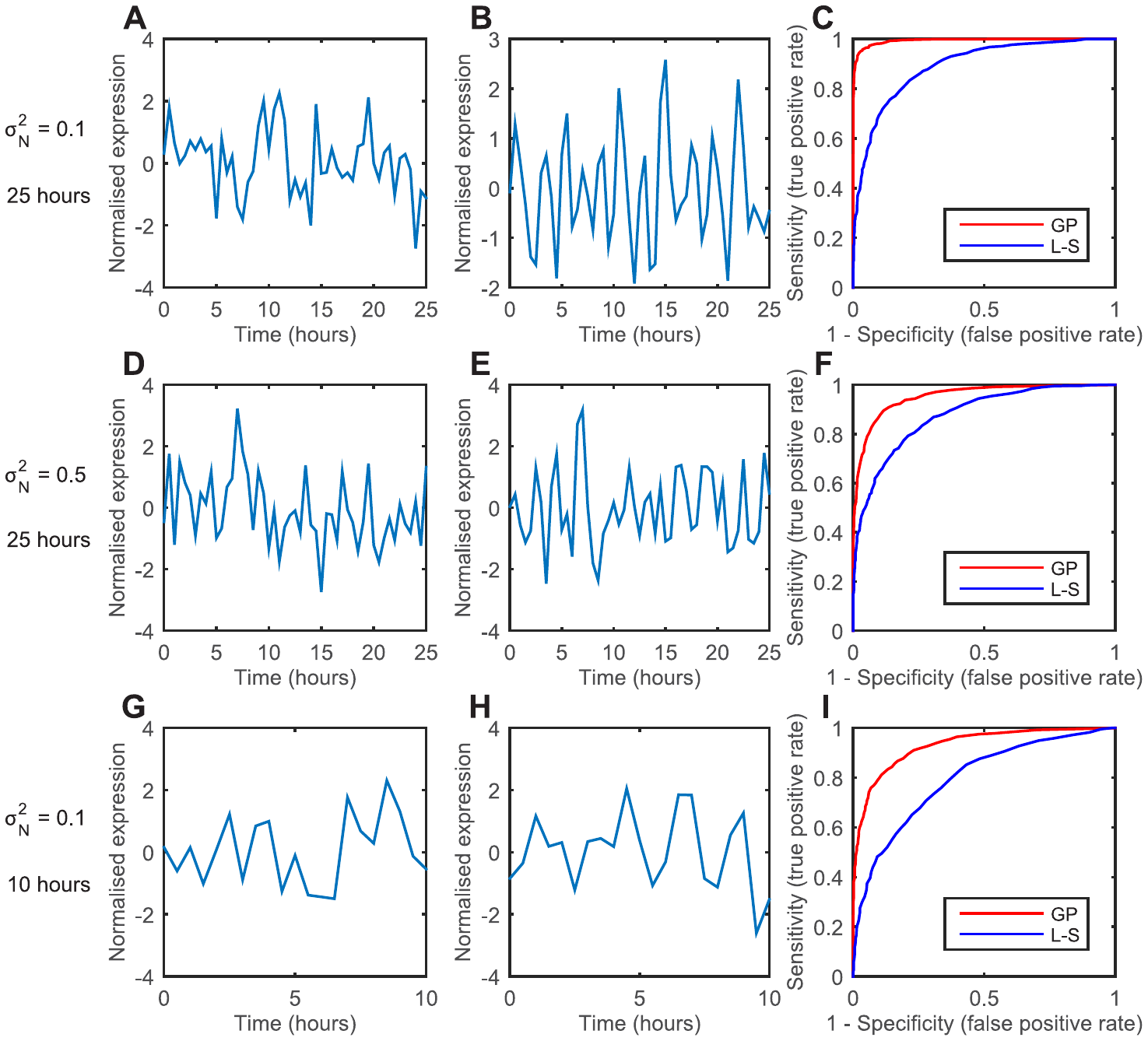}
\end{center}
\caption{{\bf Comparing the LSP and Gaussian process method using ROC curves from synthetic data.} (A, D and G) Representative time series of non-oscillatory gene expression at different noise levels and time lengths. (B, E and H) Representative time series of oscillatory gene expression at different noise levels and time lengths. (C, F and I) ROC curves: the true positive against false positive rate at different noise levels and time lengths. Red, LLR Gaussian process method; blue, LSP. }
\label{ROC}
\end{figure}

For the first synthetic data set used to generate a ROC curve, the protein levels were measured every 30 minutes for 25 hours and measurement noise with a variance of 10\% of the signal was then added to each time point (representative time series of non-oscillating and oscillating cell shown in Fig \ref{ROC}A and B, respectively). The area enclosed by the ROC curve of the new method is greater than the LSP (Fig \ref{ROC}C), and this shows that it is able to distinguish oscillating and non-oscillating cells with higher performance for the synthetic data set. When the level of experimental noise is increased to 50\% of the signal, the time series look noisier (Fig \ref{ROC}D and E). The area under the ROC curve is reduced for both methods, although the Gaussian process method still performs better (Fig \ref{ROC}F). For the final data set we consider experimental noise of 10\% of the signal and a reduced length of 10 hours (Fig \ref{ROC}G and H). The area under the curve is smaller for both methods, which demonstrates that the performance of detecting periodic signals is poorer when the time series is short, although the Gaussian process again still performs better (Fig \ref{ROC}I). Note that while the area under the ROC curve is larger for the Gaussian process method, this in isolation does not imply that statistical power is better than the LSP method. The method of FDR control determines the rejection threshold and hence position on the curve, which can in principle be more conservative with fewer true positives.

\subsection*{Effect of detrending on FDR performance}

\begin{figure}[!h]
\begin{center}
\includegraphics[scale=1]{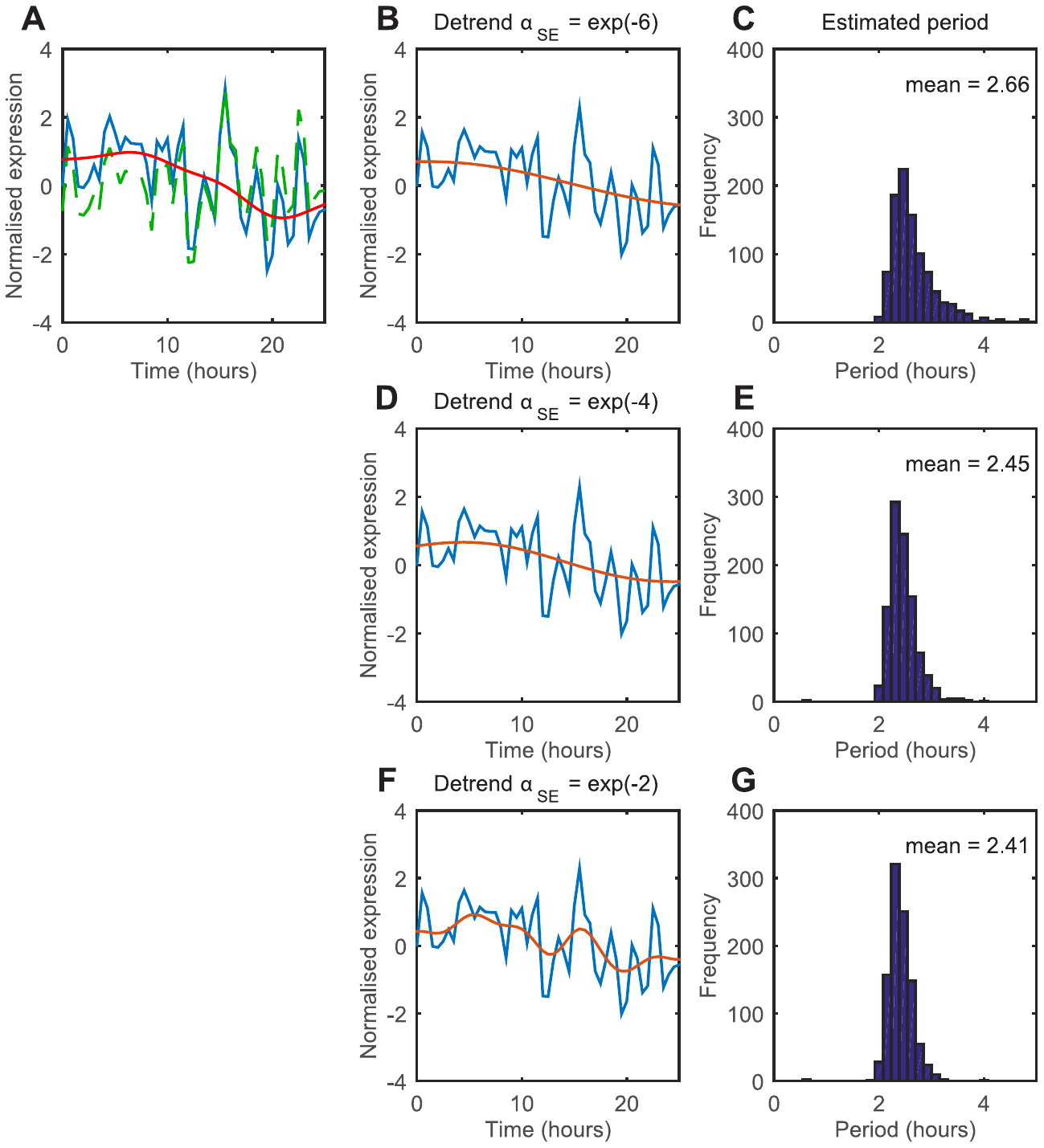}
\end{center}
\caption{{\bf The detrending length scale affects the classification of cells and the estimated period.} (A) Representative time series of oscillatory gene expression with trend added. Green, raw expression; red, added trend; blue, the sum of the two signals. (B, D and F) The fitted trend line using upper bound of $\alpha_{SE}=\exp(-6)$, $\exp(-4)$ and $\exp(-2)$, respectively. Blue, signal; red, fitted trend. (C, E and G) The estimated period of the detrended time series.}
\label{UpperBound}
\end{figure}

In order to investigate the effect of detrending on the performance of classification and period estimation, we added a long term trend at a length scale of $\alpha_{SE}=\exp(-4)$ to oscillatory and non-oscillatory synthetic data (representative time series of oscillating cell shown in Fig \ref{UpperBound}A). Each cell was then detrended using an upper bound on the detrending parameter $\alpha_{SE}$ of $\exp(-6)$, $\exp(-4)$ and $\exp(-2)$. The higher the upper bound, the more flexible the fitted trend is to the data (Fig \ref{UpperBound}B, D and F). We sought to characterise the effects of detrending on the performance of FDR estimation. The FDR estimation involves using the estimated number of non-oscillating cells, $\pi_0$, which changes for the different length scales (\ref{Pi0estimate}). When the detrending parameter is lower than the true trend and does not provide enough flexibility ($\exp(-6)$, Fig \ref{UpperBound}B), (20/1000) non-oscillating cells passed the test while (817/1000) oscillating cells passed, giving an achieved FDR of 2.4\% and true positive rate (TPR) of 82\%. When the correct length scale of $\exp(-4)$ is used (Fig \ref{UpperBound}D), (59/1000) non-oscillating cells passed the test while (955/1000) oscillating cells passed, giving an achieved FDR of 5.8\% and TPR of 96\%. Finally, when the detrending parameter provides too much flexibility ($\exp(-2)$, Fig \ref{UpperBound}F), (68/1000) non-oscillating cells passed the test while (958/1000) oscillating cells passed, giving an achieved FDR of 6.6\% and statistical power of 96\%. The average estimated period of the oscillating cells is also affected by the detrending length scale, where the period decreases from 2.66 to 2.41 hours as $\alpha_{SE}$ is increased from $\exp(-6)$ to $\exp(-2)$ (Fig \ref{UpperBound}C, E and G).

\subsection*{Calibrating the detrending parameter}

We sought to quantify the effect of the lengthscale in a systematic way to provide a guideline for a sensible parameter choice to achieve good statistical power while controlling the FDR at a low level. The detrending pipeline is designed to remove long term trends within the data, but if the lengthscale is too short then it can remove the oscillatory signal. We therefore tested the effect of the detrending lengthscale on the false positive rate, statistical power and FDR when a range of different trends is added to synthetic data from the {\it Hes1} model.

When no trend is added to synthetic data but the lengthscale of detrending is very short, the oscillations are removed and statistical power is lost (middle panel, \ref{EffectDetrending} A). Conversely, if the detrending lengthscale is too long then statistical power is lost in synthetic cells with an added trend, as the detrending lengthscale is too slow to remove the trend. This is observed over a range of added trends, using a trend parameter $\alpha_{SE}$ of $\exp(-3.5)$, $\exp(-4)$ and $\exp(-4.5)$, corresponding to a lengthscale of 4.1, 5.2 and 6.7 hours, respectively (middle panel, \ref{EffectDetrending} B-D). At a detrending lengthscale of approximately 3 times the oscillation period (7.5 hours) there is a good balance between statistical power (82-97\%) and FDR (5.1-7.2\%). We therefore recommend a default setting of roughly 3 times the expected period, but this can be adjusted within the package.

To investigate if this recommended setting generalises to other systems we also simulated a model of p53 dynamics. The p53-mdm2 circuit responds to DNA double-stranded breaks induced by gamma irradiation \cite{Geva-Zatorsky2010}. Double-stranded DNA breaks activate p53, which in turn transcriptionally activates mdm2. Mdm2 subsequently targets p53 for degradation, and hence the p53-mdm2 network forms a negative feedback loop. After gamma irradiation the levels of p53 oscillate with a period of around 6 hours. We simulated a model of p53 dynamics that has been demonstrated to well-describe the observed oscillatory dynamics (Model IV, \cite{Geva-Zatorsky2006}) and used parameters that were found to be the best fit to data (\ref{p53} A). To simulate non-oscillatory dynamics we removed the negative feedback by making p53 degradation independent of Mdm2 (\ref{p53} B). We then also quantified the effect of the detrending lengthscale by adding trends at $\alpha_{SE}$ of $\exp(-5)$ and $\exp(-6)$, corresponding to a lengthscale of 8.6 and 14.2 hours, respectively (\ref{p53} C, D). Using the recommended detrending lengthscale of 18 hours controls the FDR to a reasonable level of around 6.6\% in both cases.

Note that the reason why the achieved FDR can exceed the expected value of 5\% is due to approximations used in our parametric bootstrap approach. Approximations enter at various steps, and the synthetic bootstrap data does not perfectly represent the Gillespie data used to simulate cells, which leads to an LLR distribution of non-oscillating cells in the Gillespie simulations that is broader than the synthetic bootstrap distribution (\ref{SuppFigGillespieWithTrendOffset}). The difference between the true non-oscillating LLR distribution and the bootstrap approximation can be quantified with the Kolmogorov-Smirnov (KS) distance. The KS distance using the bootstrap approach (0.066, \ref{CompareChi2} A) is smaller than than that obtained with a Chi-squared distribution with one degree of freedom to account for extra parameter (0.21, \ref{CompareChi2} B), and hence the bootstrap method is a better proxy for the null distribution, although it is not perfect. The first challenge is to perfectly disentangle trends from the underlying signal, and the variance of the trend may be underestimated. The second is that Gillespie simulations are only approximated by Gaussian processes, and even without any trend added the LLR distribution do not perfectly match (\ref{SuppFigGillespieNoTrend}). However, even when non-oscillating data are generated by an OU Gaussian process the average LLR score is still lower in the synthetic bootstrap (\ref{SuppFigOUnotrend}). This may be because the time scale of fluctuations is typically underestimated i.e. fluctuations are inferred to be more rapid than they really are. This issue of biased parameter estimation is mitigated when the time series is longer and the averages are better matched. 

In conclusion, the method of classifying cells based on Gaussian processes is more effective than the LSP on data simulated with a stochastic model over a range of noise levels and time lengths. Our estimated FDR provides a useful threshold for controlling false discoveries and we suggest a detrending lengthscale to provide a compromise between false discoveries and statistical power. 

\subsection*{Model mis-specification}

During the derivation of our method there are several approximations to take a forward model of a general stochastic network and represent it as a simple Gaussian process with one of two covariance functions. When the underlying dynamics of the network are not well-captured by our approximation then our model may become mis-specified. To test the robustness of our method to model mis-specification, we considered three ways this can occur:

1) The LNA is valid at large system size, so when the number of molecules is low (low system size) then the LNA can lose accuracy and the system may no longer behave like a Gaussian process. These effects can be particularly strong when there are nonlinear reaction steps in the system, the parameters of the model lead to oscillations (near a Hopf bifurcation) and the copy number of interacting species is low \cite{Thomas2013}. To validate that our approach still worked on non-Gaussian data we simulated the {\it Hes1} network at a low system size of $\Omega$ = 1. Here, the waveform becomes asymmetric (\ref{LowSystemSize} A), and the magnitude of peaks are larger than troughs. This also leads to an asymmetry in a histogram of all the data points (\ref{LowSystemSize} B), with a skew towards larger values. Our method still works well in this regime and successfully classified 99\% of oscillating cells, despite the underlying data deviating from a Gaussian process.

2) The LNA works by considering fluctuations around a single steady state. Some networks are able to generate bistability, with two steady states. An example of such a network is shown in (\ref{Bistable} A), where two transcription factors mutually repress each other’s transcription and lead to stable protein 1 high/ protein 2 low (or vice versa) configurations. Fluctuations caused by intrinsic noise are able to switch the system from one stable configuration to the other, as shown in examples (\ref{Bistable} B and \ref{Bistable} C). Occasionally the switching can look rhythmic, leading to a high LLR score (\ref{Bistable} B), whilst other times the score is low (\ref{Bistable} C). The behaviour of a bistable network at a population level was simulated with 2000 cells, and the LLR score is shown in \ref{Bistable} D. The LLR score produced by a non-oscillating population of OU bootstrap samples shows a very similar distribution (\ref{Bistable} E). In this case the test becomes very conservative, and no cells from the bistable network pass as oscillators.

3) During the derivation we assume there is only one dominant behaviour within the time series (one eigenvalue of the covariance function, and hence only one frequency). This assumption can be violated when the dynamics are more complicated, and an example of this is when there are two characteristic frequencies within the data. \ref{TwoFrequencies} A shows a time series created by adding two OUosc functions with different frequencies together: a 2.5 and a 24 hour period. If the expected period of 2.5 hours is used to detrend with a length scale of 7.5 hours (3x expected), then the trend matches the long period oscillation (red, \ref{TwoFrequencies} A). When this trend is removed then the short period oscillations are recovered (\ref{TwoFrequencies} B), but the longer frequency is filtered out. 

Taken together, these examples show that our method is robust to at least some common cases of mis-specification, although manual inspection of the data is recommended to establish whether the time series data differ substantially from an OU or OUosc Gaussian process. The data analysis pipeline can be in principle be adapted to handle more complicated dynamics through adding multiple covariance functions together and including methods such as changepoint detection (see Discussion). 
\subsection*{Application to bioluminescence imaging data}

\begin{figure}[!h]
\begin{center}
\includegraphics[scale=1]{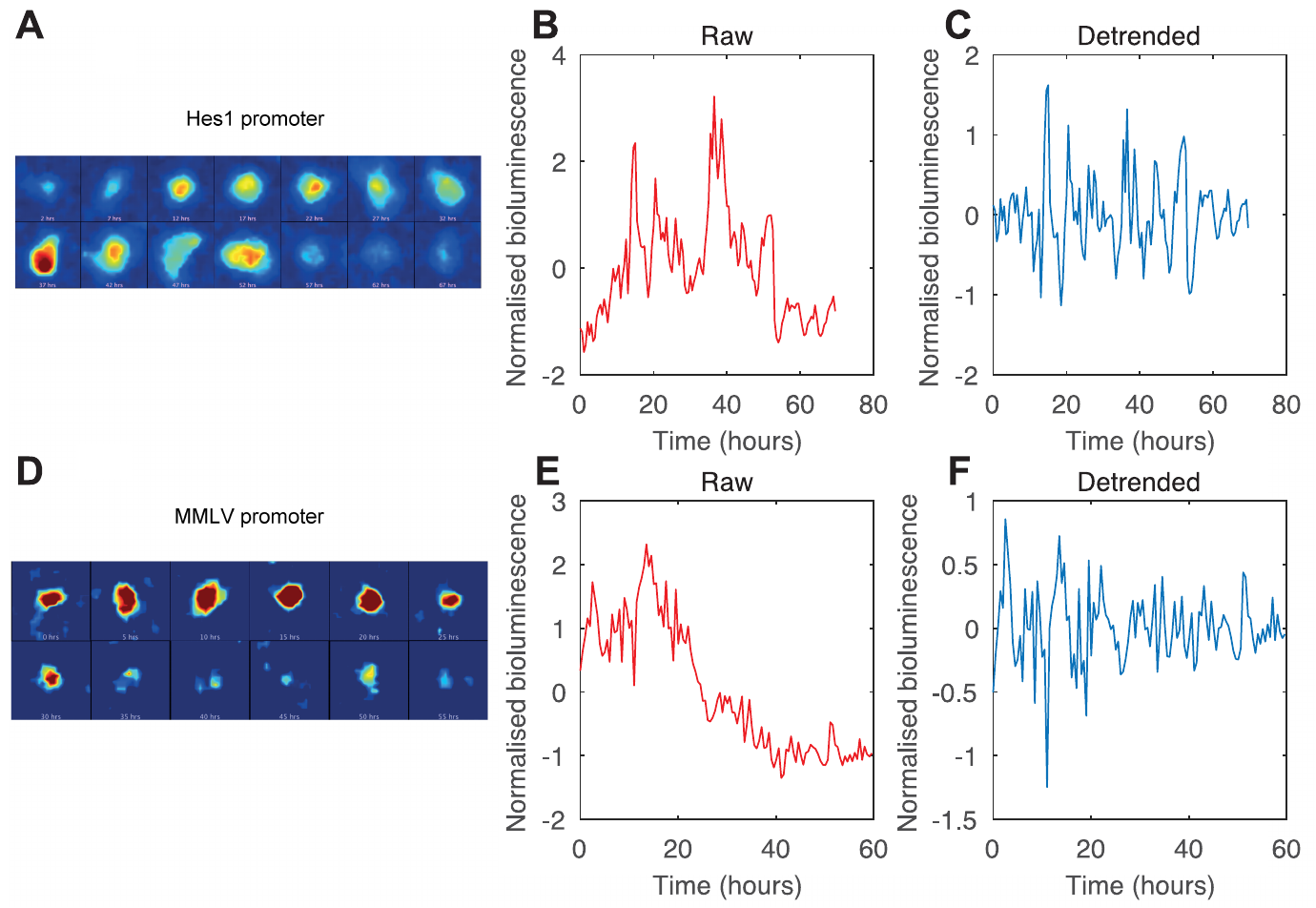}
\end{center}
\caption{{\bf Bioluminescence imaging of C17 neural progenitor cells with luciferase expression driven by the oscillatory {\it Hes1} or constitutive MMLV promoter.} (A) Bioluminescence images of an individual cell with luciferase driven by {\it Hes1} promoter. (B) Raw time series of bioluminescence quantification from A. (C) Detrended bioluminescence time series. (D) Bioluminescence images of an individual cell with luciferase driven by MMLV promoter. (E) Raw time series of bioluminescence quantification from D. (F) Detrended bioluminescence time series. Both reporters show dynamic expression.}
\label{Images}
\end{figure}

Having demonstrated the utility and accuracy of our method on synthetic data, we then applied the method to an experimental data set consisting of time series from live single-cell bioluminescence imaging. A firefly luciferase reporter was driven either by the 2.7 Kb region of the {\it Hes1} promoter or by a constitutive MMLV promoter. The bioluminescence of single cells was measured by exposing the camera to signal over 30 minutes, to give time points every 30 minutes, where the length of time series ranged from 26 to 72 hours. The {\it Hes1} promoter has been previously reported to drive oscillatory expression due to negative feedback of the HES1 protein on its own promoter \cite{Masamizu2006, Shimojo2008}, although the proportion of oscillating cells within an experimental population of cells has not been previously reported. Examples of the time series generated by a sequence of bioluminescent images from both promoters are shown in Fig \ref{Images}. As can be seen in the traces of single cells in Fig.~\ref{Images}, both the oscillatory (\ref{Images}A-C) and the constitutive reporter (\ref{Images}D-F) show dynamic expression with many peaks and troughs. Intuition is insufficient to classify these cells, and an objective statistical measure must be used.

\begin{figure}[!h]
\begin{center}
\includegraphics{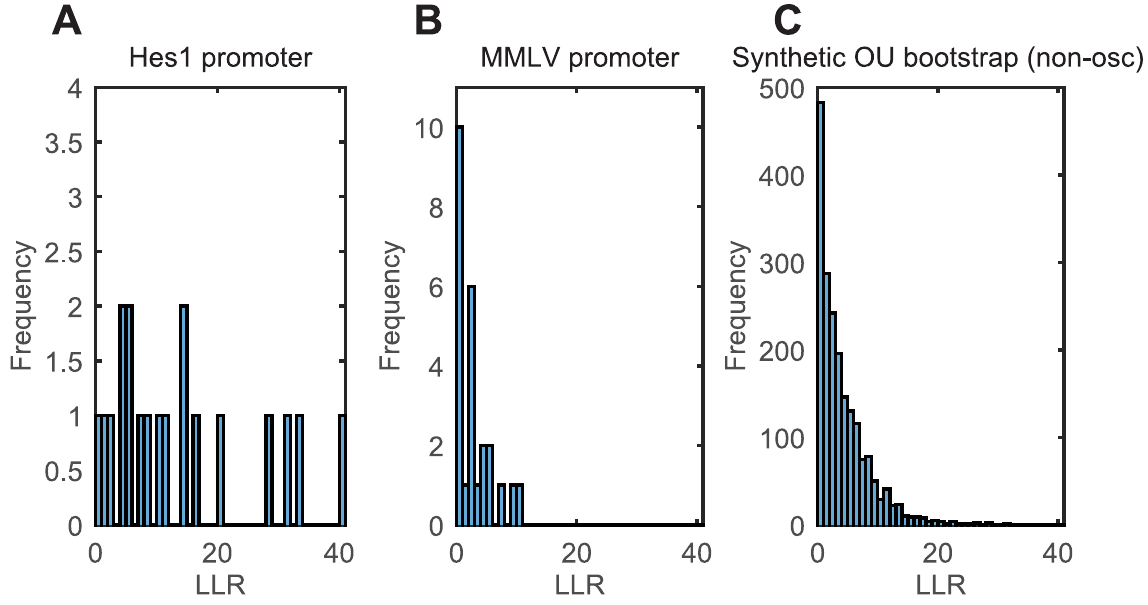}
\end{center}
\caption{{\bf The LLR between OU and OUosc model fits to the single-cell experimental data and synthetic non-oscillatory data.} (A) The LLR distribution of 19 cells with the {\it Hes1} promoter. (B) The LLR distribution of 25 cells with the MMLV promoter. (C) The LLR distribution of 2000 synthetic cells generated from the OU model using parameters fitted to A and B.}
\label{LLRdists}
\end{figure}

\begin{figure}[!h]
\begin{center}
\includegraphics{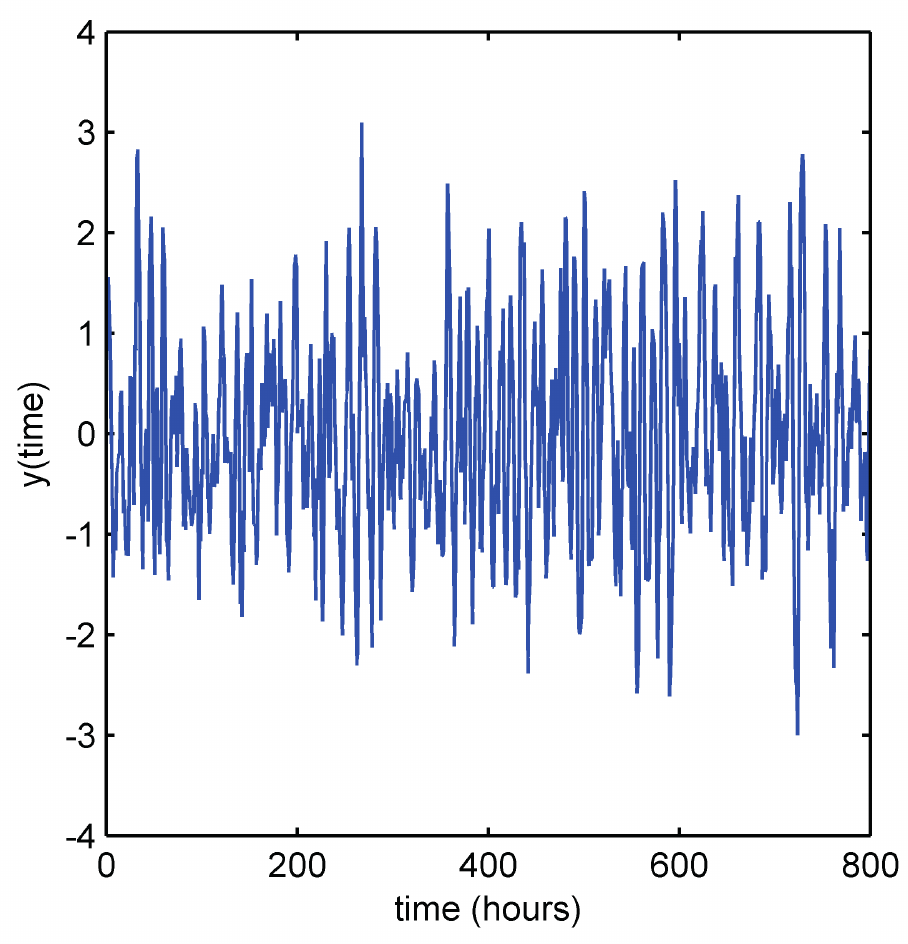}
\end{center}
\caption{Long simulated time series examples from the OUosc covariance functions. OUosc parameter values are $\alpha = 0.2$, $\beta = 0.5$, and $\sigma = 1$.}
\label{LongTimeSeriesExample2}
\end{figure}

\begin{figure}[!h]
\begin{center}
\includegraphics{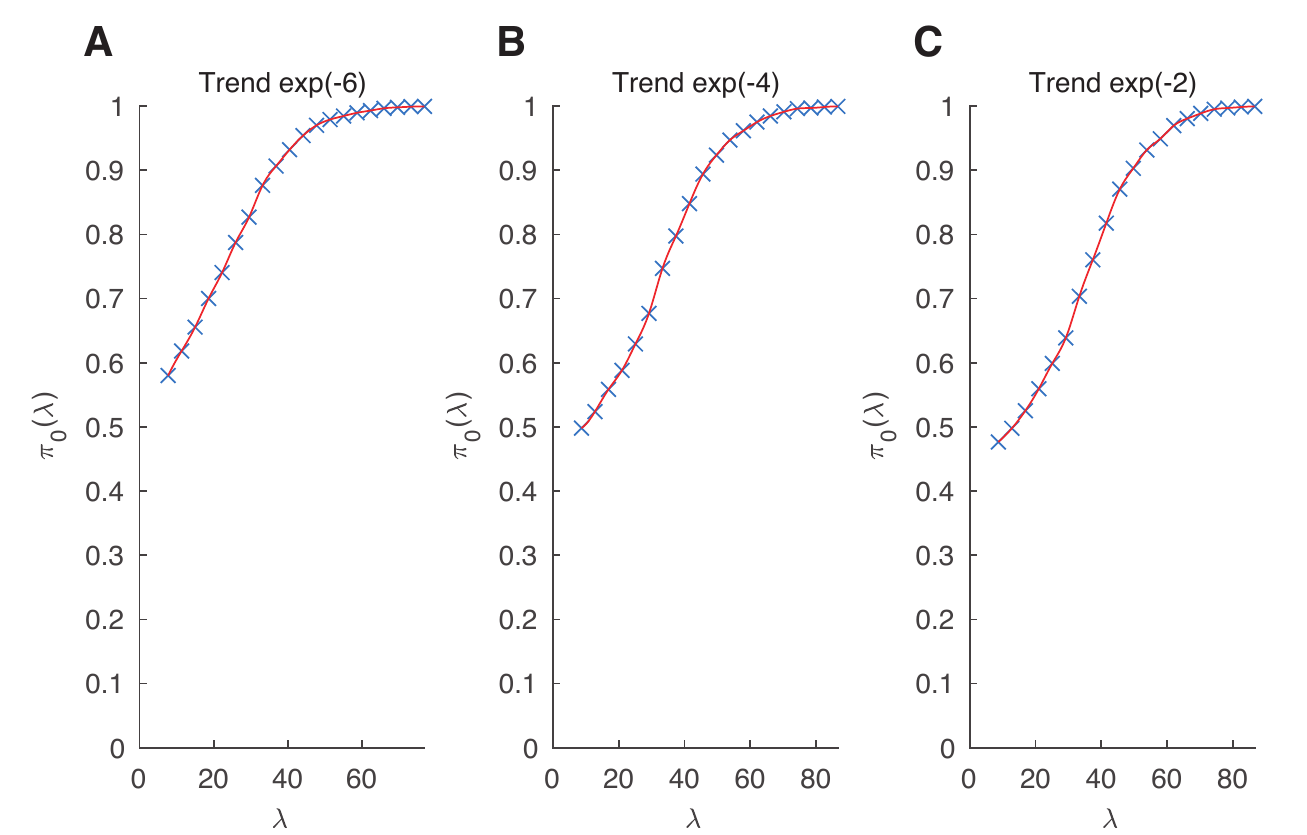}
\end{center}
\caption{The effect of the detrending length scale on the estimated number of non-oscillating cells ($\pi_0$). (A, B and C) The estimated number of non-oscillating cells, $\pi_0$, using a detrending length scale of  $\alpha_{SE}=\exp(-6)$, $\alpha_{SE}=\exp(-4)$ and $\alpha_{SE}=\exp(-2)$, respectively. The value of $\pi_0$ is estimated as the spline fit (red line) at the lowest $\lambda$.}
\label{Pi0estimate}
\end{figure}

\begin{figure}[!h]
\begin{center}
\includegraphics[scale=0.9]{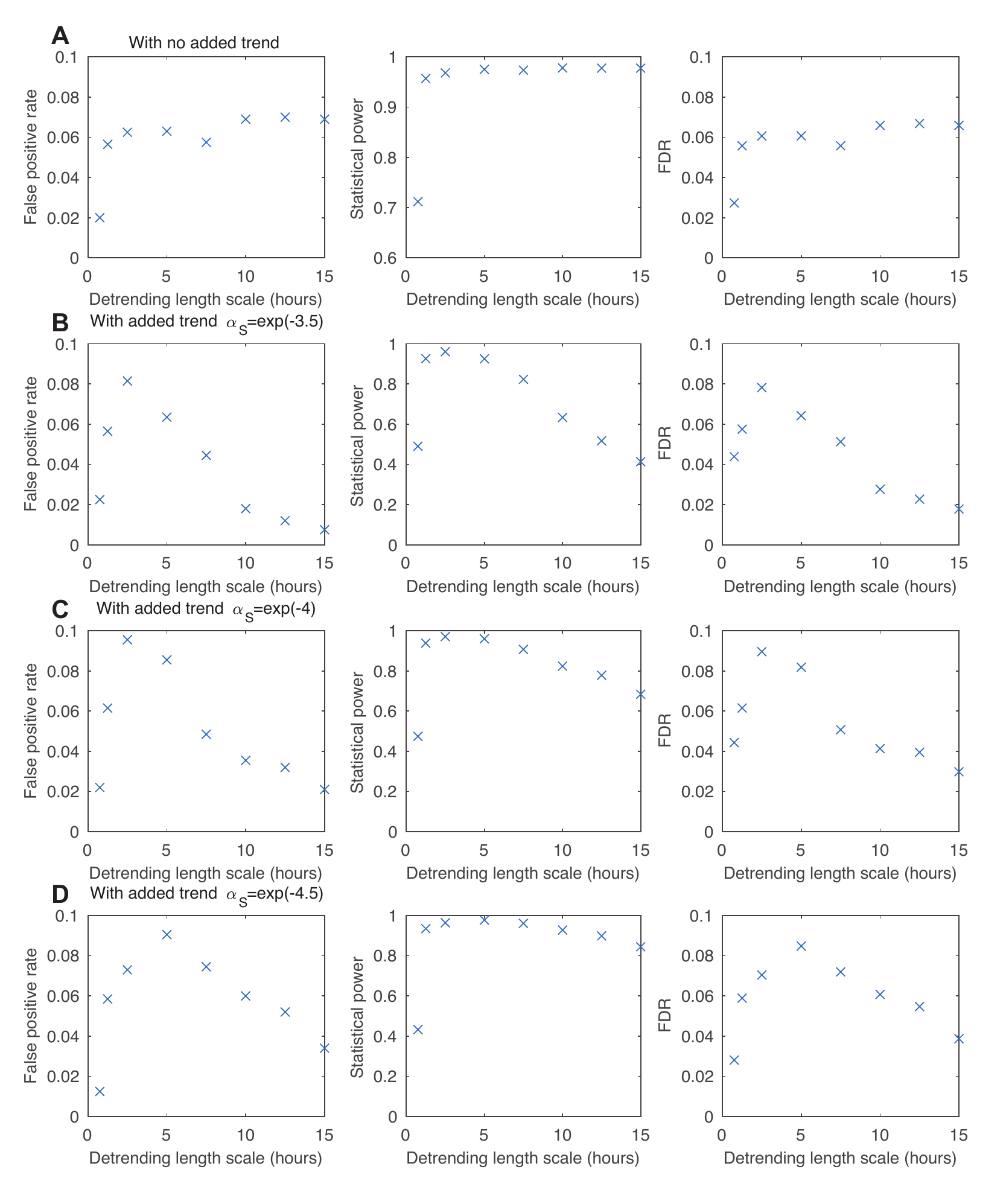}
\end{center}
\caption{The false positive rate, statistical power and FDR of 2000 oscillating and non-oscillating cells simulated with the Gillespie algorithm. (A) with no trend added, trend added at (B) $\alpha_{SE}=\exp(-3.5)$, (C) $\alpha_{SE}=\exp(-4)$ and (D) $\alpha_{SE}=\exp(-4.5)$.}
\label{EffectDetrending}
\end{figure}

\begin{figure}[!h]
\begin{center}
\includegraphics[scale=0.9]{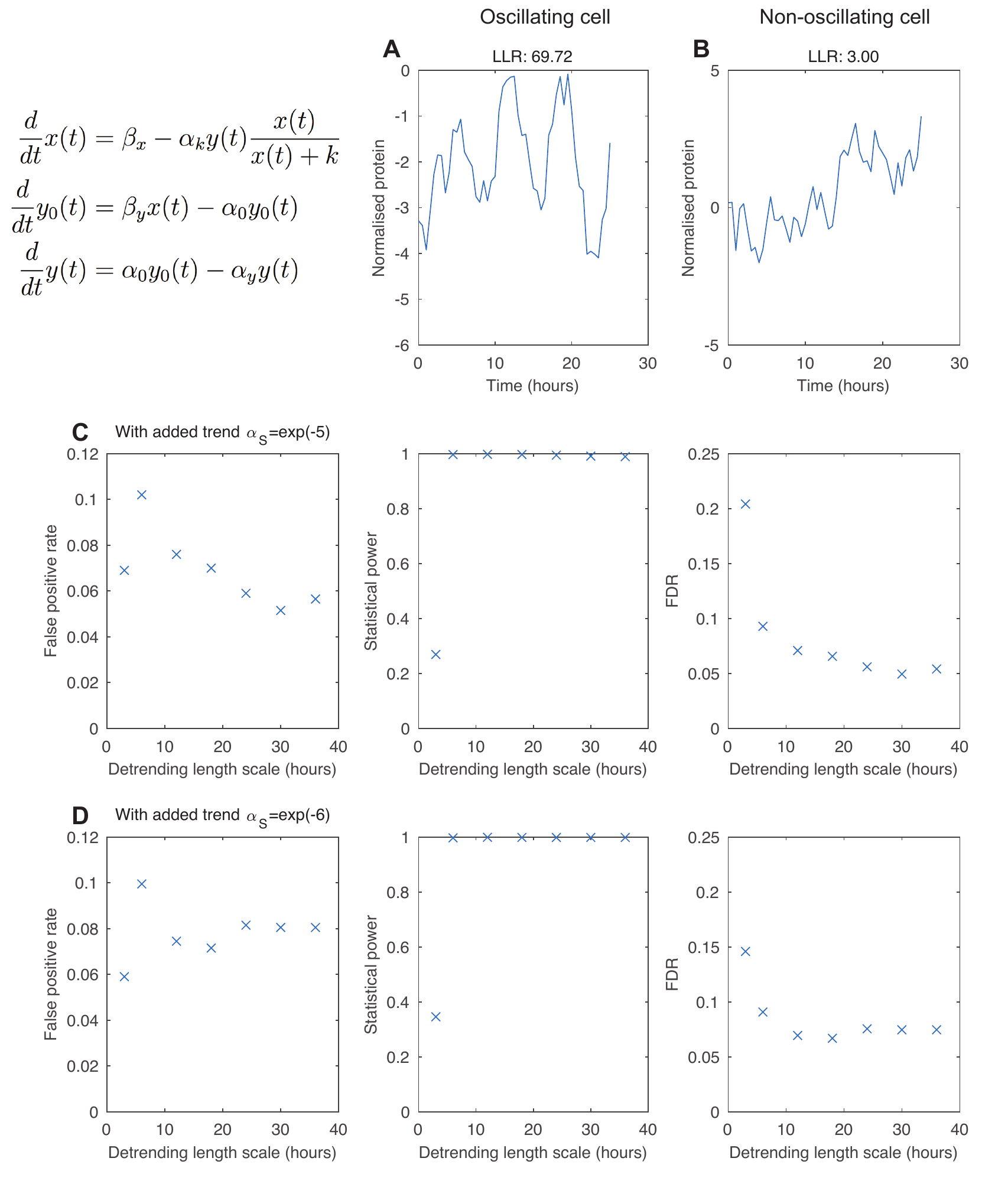}
\end{center}
\caption{Assessing the effect of detrending using a model of p53 dynamics. (A) Time series example for the p53 model in the oscillatory regime. Model parameters are $\beta_x$ = 0.9,  $\alpha_k$ = 1.7, $k$ = 0.0001, $\beta_y$ = 1.1, $\alpha_0$ = 0.8, $\alpha_y$ = 0.8 and $\Omega$ = 20. (B) Time series example for the p53 model in the non-oscillatory regime, where $y(t)$ is removed from the degradation of $x(t)$. Model parameters are $\beta_x$ = 0.9,  $\alpha_k$ = 5.1, $k$ = 1, $\beta_y$ = 1.1, $\alpha_0$ = 0.8, $\alpha_y$ = 0.8. and $\Omega$ = 20. (C, D) The false positive rate, statistical power and FDR of 2000 oscillating and non-oscillating cells from the p53 model simulated with the Gillespie algorithm with trend added at (C) $\alpha_{SE}=\exp(-5)$, (D) $\alpha_{SE}=\exp(-6)$.}
\label{p53}
\end{figure}

\begin{figure}[!h]
\begin{center}
\includegraphics{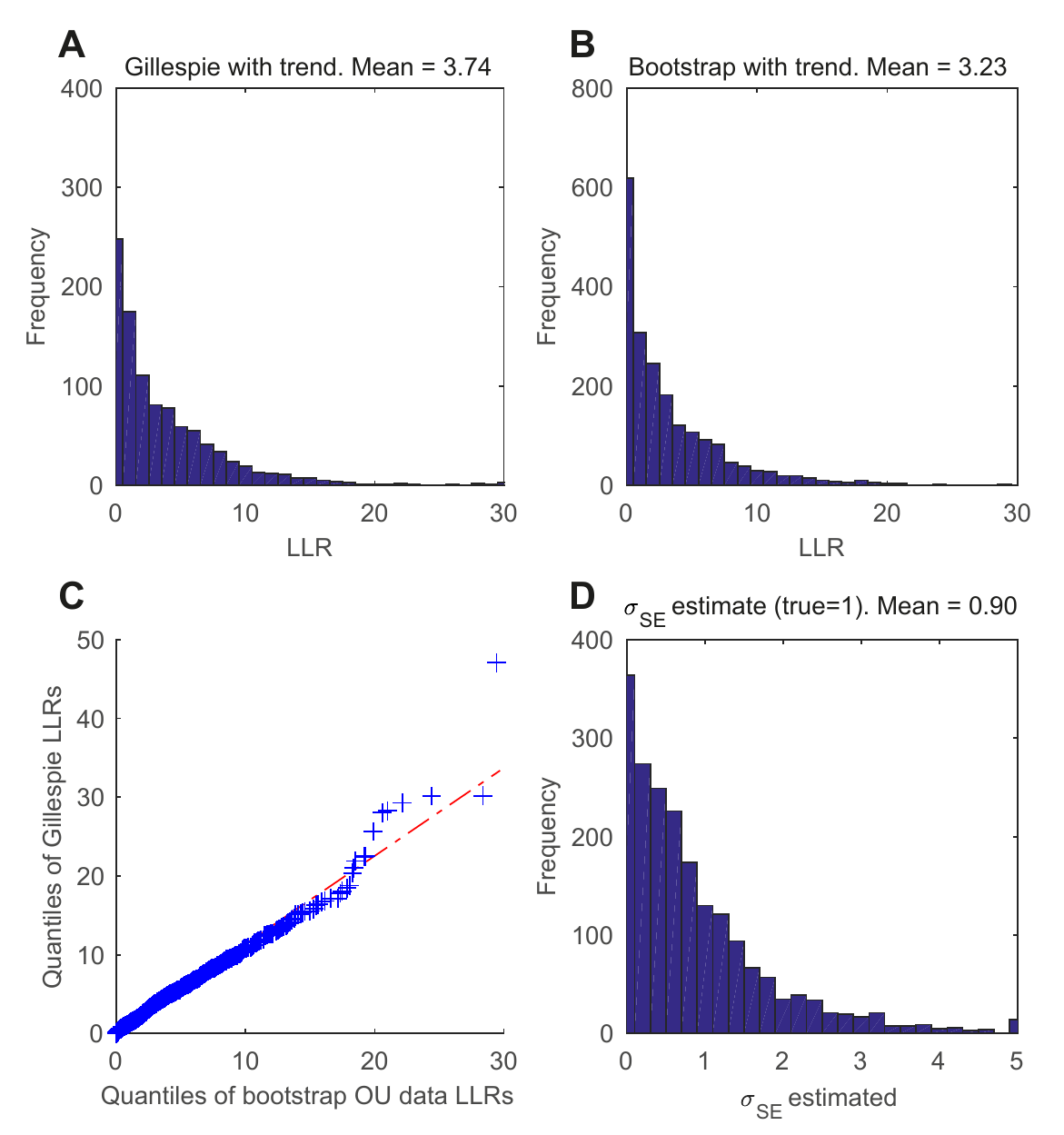}
\end{center}
\caption{Comparison of the LLR distribution generated by the non-oscillating Gillespie simulations with added trend of $\alpha_{SE}=\exp(-4)$ and the corresponding LLR distribution of the synthetic bootstrap data of the entire data set. (A) The LLR distribution of the of non-oscillating Gillespie simulations with added trend of $\alpha_{SE}=\exp(-4)$. (B) The LLR distribution of synthetic bootstrap data of the entire data set. (C) The Q-Q plot of the Gillespie simulated (plus trend) LLR distribution (from A) against the OU bootstrap LLR distribution (B). (D) The estimates of $\sigma_{SE}$ inferred from the Gillespie data with trend added (true value is 1).}
\label{SuppFigGillespieWithTrendOffset}
\end{figure}

\begin{figure}[!h]
\begin{center}
\includegraphics{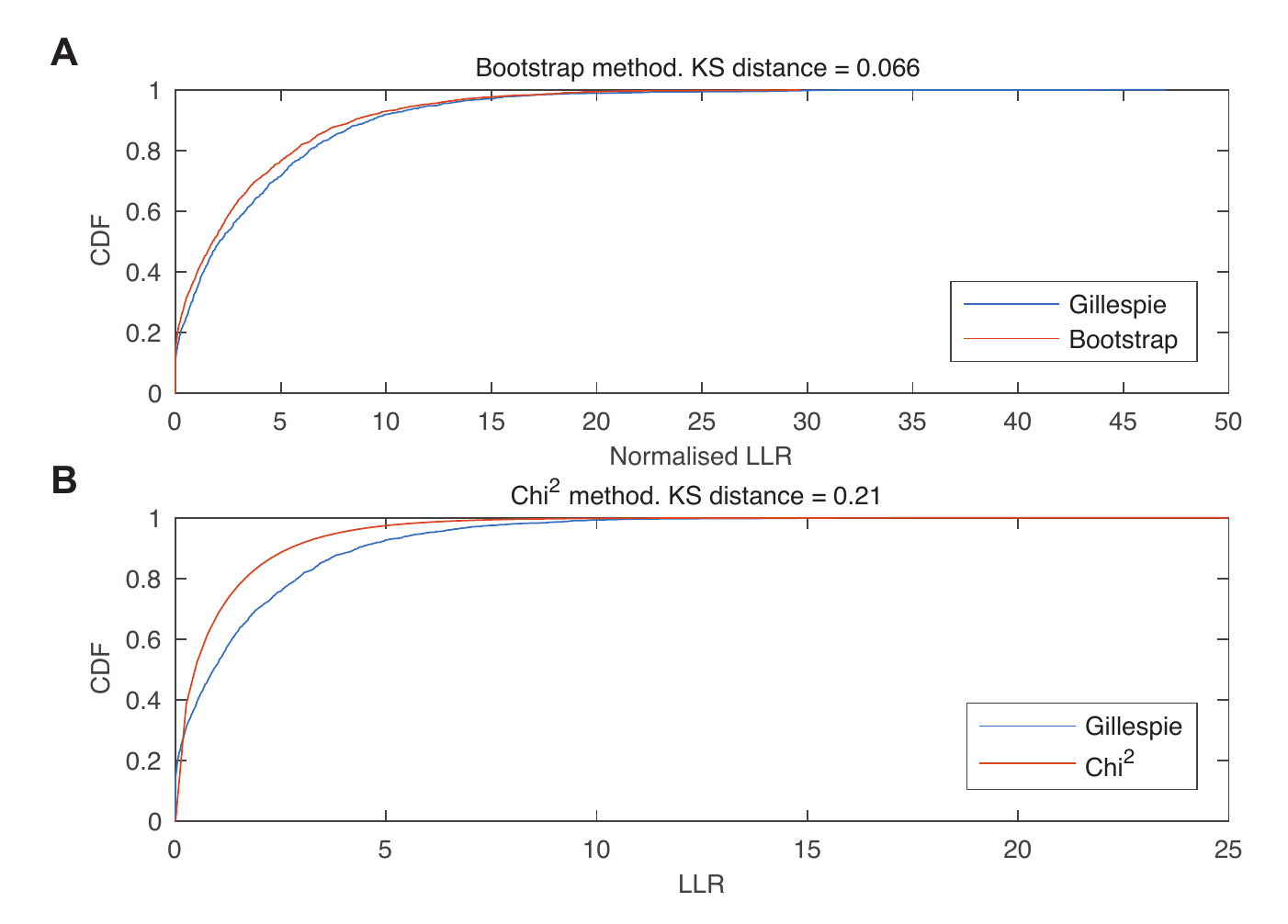}
\end{center}
\caption{Comparing the LLR distribution of non-oscillating Gillespie simulations with synthetic bootstrap and chi-squared distributions. (A) The cumulative density function of the LLR of 1000 non-oscillating Gillespie simulations with added trend of $\alpha_{SE}=\exp(-4)$ (from \ref{SuppFigGillespieWithTrendOffset} A) and the corresponding LLR distribution of the synthetic bootstrap data (from \ref{SuppFigGillespieWithTrendOffset} B). Note that LLR is normalised to the length of the data and multiplied by 100, as described in text. (B) The cumulative density function of the LLR of 1000 non-oscillating Gillespie simulations with added trend of $\alpha_{SE}=\exp(-4)$ (from \ref{SuppFigGillespieWithTrendOffset} A) and the chi-squared distribution with one degree of freedom. The LLR is not normalised.}
\label{CompareChi2}
\end{figure}

\begin{figure}[!h]
\begin{center}
\includegraphics{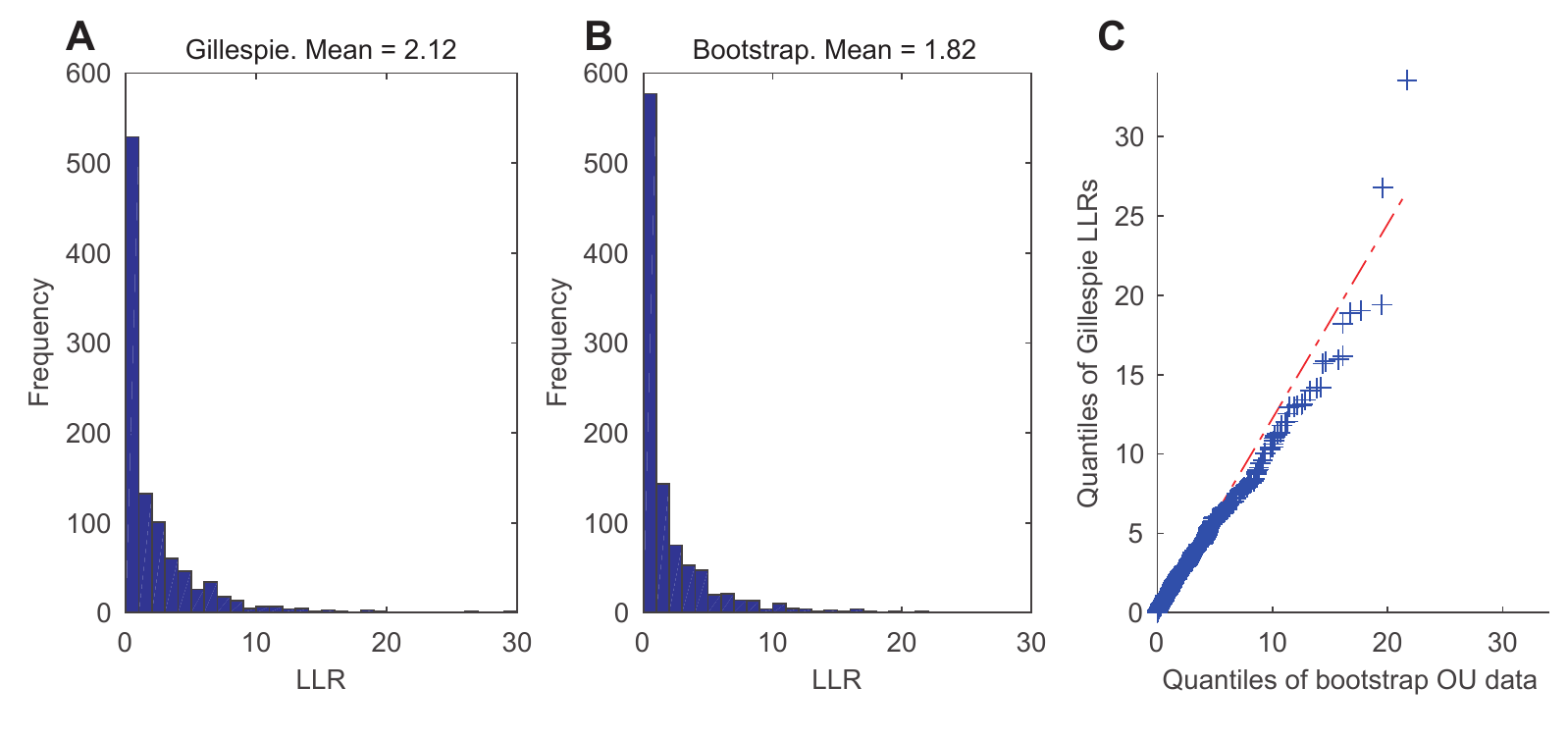}
\end{center}
\caption{Comparison of the LLR distribution generated by the non-oscillating Gillespie simulations with no added trend and the corresponding LLR distribution of the synthetic bootstrap data of the entire data set. (A) The LLR distribution of the of non-oscillating Gillespie simulations with no added trend. (B) The LLR distribution of synthetic bootstrap data of the entire data set. (C) The Q-Q plot of the Gillespie simulation LLR distribution (from A) against the OU bootstrap LLR distribution (B).}
\label{SuppFigGillespieNoTrend}
\end{figure}

\begin{figure}[!h]
\begin{center}
\includegraphics[scale=0.9]{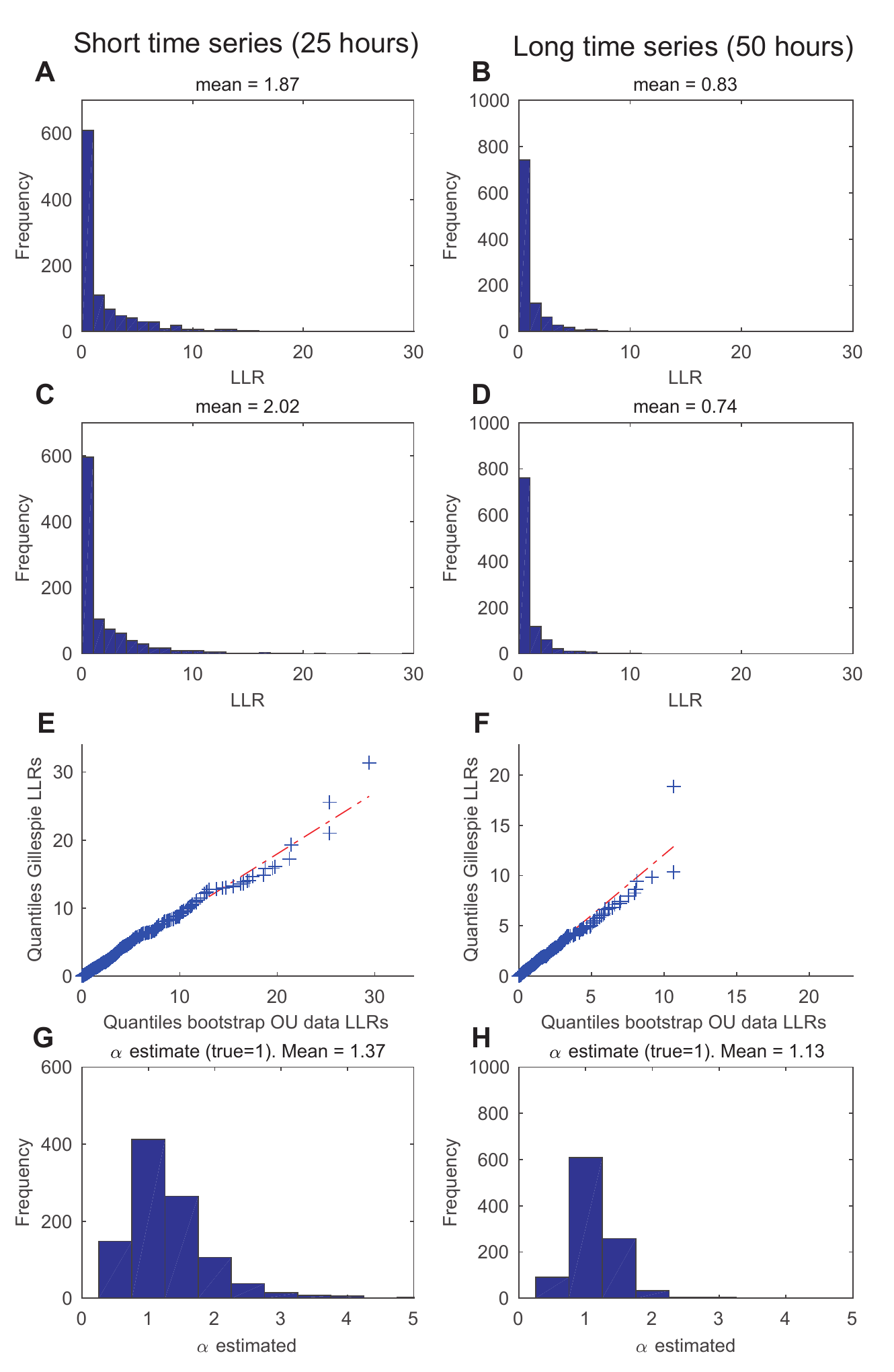}
\end{center}
\caption{Comparison of the LLR distribution generated by an OU Gaussian process ($\alpha = 1$ and $\sigma = 1$) with no added trend and the corresponding LLR distribution of the synthetic bootstrap data of the entire data set. (A, B) The LLR distribution of the of $\alpha_{SE}=\exp(-4)$ for time lengths of 25 and 50 hours, respectively. (C, D) The LLR distribution of synthetic bootstrap data of the entire data set for time lengths of 25 and 50 hours, respectively. (E, F) The Q-Q plots of the OU simulated LLR distribution against the OU bootstrap LLR distribution for time lengths of 25 and 50 hours, respectively. (G, H) The estimates of $\sigma_{SE}$ in from the Gillespie data (true value is 1) for time lengths of 25 and 50 hours, respectively.}
\label{SuppFigOUnotrend}
\end{figure}

\begin{figure}[!h]
\begin{center}
\includegraphics[scale=0.9]{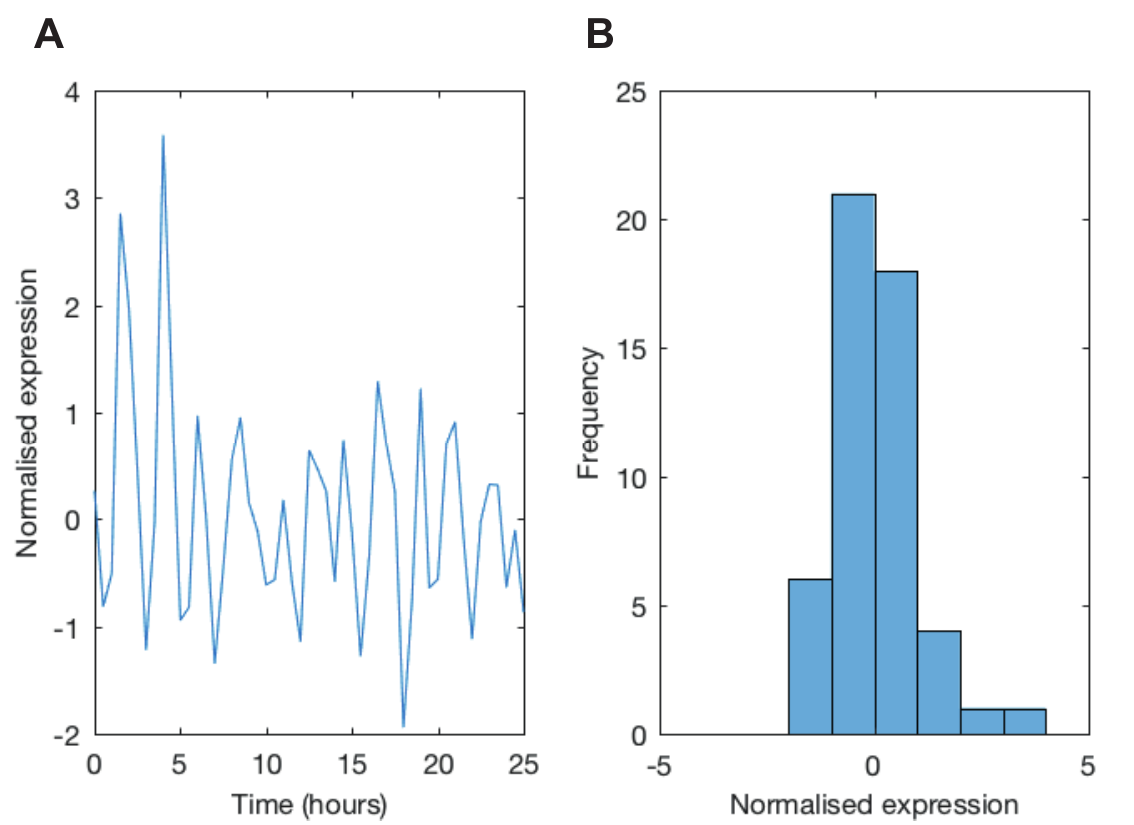}
\end{center}
\caption{Illustrative low system size simulation of the {\it Hes1} oscillator. (A) Time series example of {\it Hes1} oscillator at a system size of $\Omega = 1$. (B) Histogram of all data points contained in (A).}
\label{LowSystemSize}
\end{figure}

\begin{figure}[!h]
\begin{center}
\includegraphics{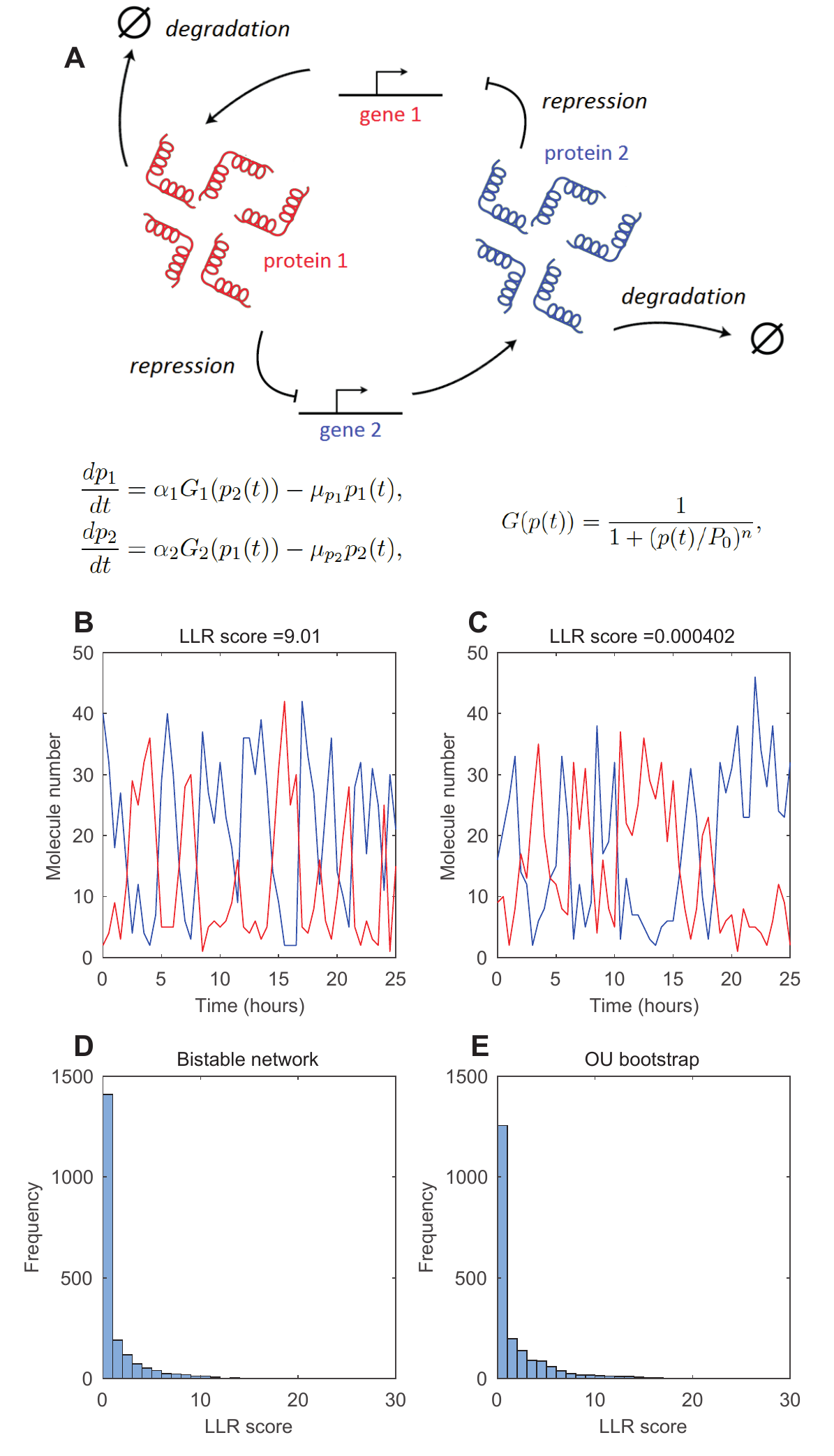}
\end{center}
\caption{Assessing the method performance on a bistable network. (A) Network topology of the bistable network. (B, C) Time series examples of bistable network. Model parameters are $P_0$ = 12, $n$ = 2, $\alpha_m$ = $\alpha_p$ = 10, $\mu_m$ = $\mu_p$ = 0.3 and $\Omega$ = 1. (D, E) LLR distributions of 2000 cells simulated from bistable network and from OU bootstrap, respectively.}
\label{Bistable}
\end{figure}

\begin{figure}[!h]
\begin{center}
\includegraphics{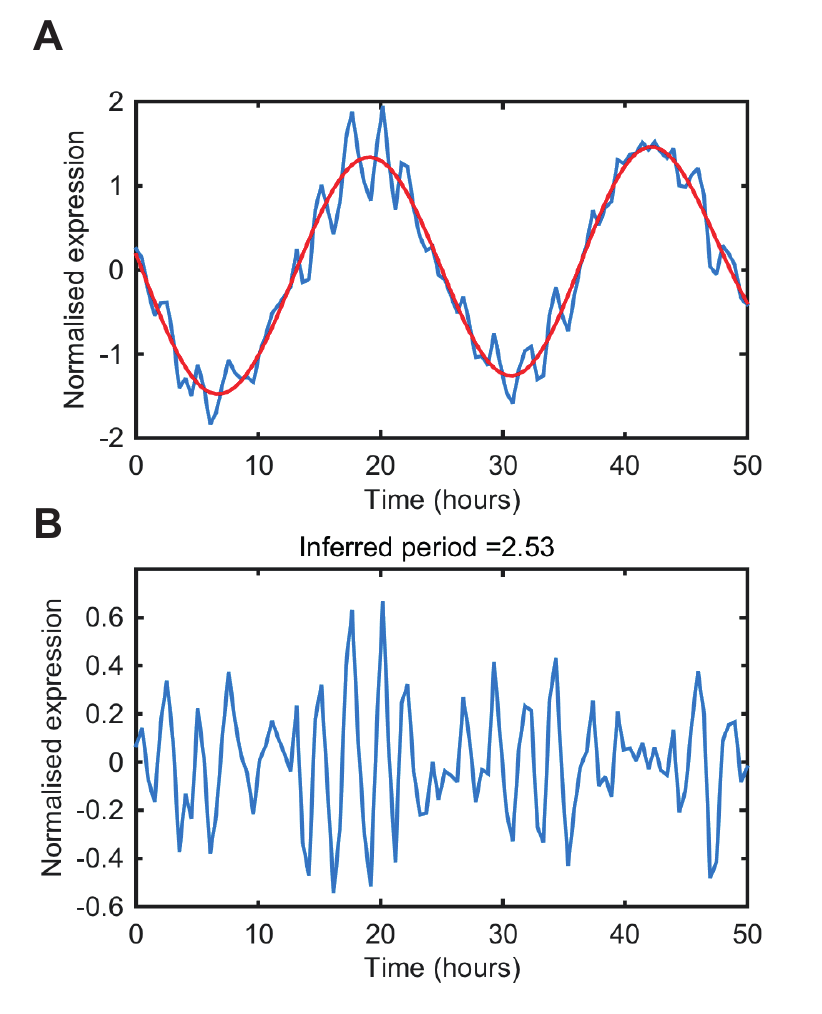}
\end{center}
\caption{Assessing the method performance on time series containing two frequencies. (A) Time series example of dynamics generated by two oscillatory OUosc covariance functions added together, with a period of 2.5 and 24 hours. Covariance parameters are: $\sigma_1$ = 5, $\alpha_1$ = 0.001, $\beta_1 = 2\pi/24$, $\sigma_2$ = 1, $\alpha_2$ = 0.1, $\beta_2 = 2\pi/2.5$. (B) The corresponding time series from (A) after detrending with a lengthscale of 7.5 hours.}
\label{TwoFrequencies}
\end{figure}

All cells from both the {\it Hes1} (19 cells) and MMLV promoter (25 cells) were analysed by calculating the LLR between OU and OUosc models. The LLRs of the {\it Hes1} promoter cells ranged from 0 to 40, with a reasonably uniform distribution over the entire range (Fig \ref{LLRdists}A). The distribution of the LLRs for the MMLV promoter cells were more peaked towards lower values, with fewer cells having a high LLR (Fig \ref{LLRdists}B). The FDR is then required to decide the LLR cut-off threshold and quantify the numbers of oscillating cells in each population.

Using the parameters of the trend and OU model fitted to both the cells with {\it Hes1} and MMLV promoter, we simulated 2000 synthetic cells from the OU model and calculated the LLR. This defined our null distribution of LLRs, representing the distribution expected from a population of non-oscillating cells (Fig \ref{LLRdists}C). The LLR distribution from the null distribution was more similar to the MMLV promoter than the {\it Hes1} promoter (compare Fig \ref{LLRdists}B with C), indicating that the proportion of non-oscillating cells in the MMLV promoter data set was higher. By comparing the null distribution with the experimental data from the {\it Hes1} and MMLV promoter we obtained an estimate of the FDR at various LLR thresholds, and we control the FDR at 5\%. At an FDR of 5\%, 10/19 of the cells from the {\it Hes1} promoter exceed the LLR threshold and are classified as oscillatory, whereas 0/25 cells from the MMLV promoter pass. This shows that the method is able to classify a higher proportion of cells as oscillatory from single cell gene expression data of a promoter with known oscillatory behaviour relative to a promoter with constitutive expression. If the FDR is made less stringent and controlled at 10\%, 12/19 {\it Hes1} and 0/25 MMLV cells pass the test.

Applying the LSP to the same data set leads to all cells passing as oscillators (19/19 {\it Hes1} and 24/25 MMLV cells pass). Using our developed detrending protocol and applying the LSP on the detrended data with Benjamin-Hochberg FDR correction as presented in \cite{Glynn2006} leads to 11/19 {\it Hes1} and 4/25 MMLV cells passing the test. This suggests that our detrending pipeline may be used by other methods to improve performance, although the FDR may still not be as good as using the OU/OUosc approach.

The various steps in the pipeline have different computational runtimes. Fitting the OU/OUosc models to the entire dataset (44 cells) take 5 minutes. Generating a synthetic population of 2000 OU non-oscillating cells takes 0.04 seconds, but re-fitting the OU/OUosc models takes 90 minutes. The LSP in contrast is faster, taking 15 seconds to analyse all cells (with detrending). The asymptotic runtime of the Gaussian process method is $O(n^3)$ for $n$ number of data points, which is limited by the Cholesky factorisation used to implement the matrix inversion in calculating the log marginal likelihood (Eq~(\ref{loglik})). If time is prohibitive then faster approximate algorithms for GP inference can be used based on several approximations, and these would be especially useful for long time series. These methods are implemented in standard GP packages including the GPML package used here, but for our purposes no approximate inference was required.

Note that the detrending parameter has an effect on the overall ranking of cells. The analysis of the data was performed with a detrending parameter of $\alpha_{SE}=\exp(-4.5)$, which corresponds to a time scale of 6.7 hours. The ranked list of LLRs is shown in S1 Table, with detrending parameters of $\alpha_{SE}=\exp(-4.5)$, $\exp(-4)$ and $\exp(-5)$.

\section*{Discussion}

Oscillations are important for a wide range of biological processes, but current methods for discovering oscillatory gene expression are best suited for regular oscillators and assume white noise as a null hypothesis of non-oscillatory gene expression. We have introduced a new approach to analysing biological time series data well suited for cases where the underlying dynamics of gene expression is inherently noisy at a single-cell level. By modelling gene expression as a Gaussian process, two competing models of single cell dynamics were proposed. An OU model describes random aperiodic fluctuations, but in contrast to white noise the fluctuations are correlated over time, and this may form a more appropriate model of non-oscillating single cells for statistical testing. The OUosc model describes quasi-periodic oscillations with a gradually drifting phase causing ``dephasing" whereby an initially synchronous population of cells lose phase with one another; such peak-to-peak variability has been observed in the {\it Hes1} system \cite{Masamizu2006}. This general class of quasi-periodic covariance function, which is the product of a decaying and oscillatory function, has been described before \cite{Solin2014} but has not to our knowledge been derived from a mechanistic model of a chemical reaction system. Gaussian processes have previously been used to detect periodicity in biological time course data \cite{Durrande2016} but not in the context of quasi-periodic stochastic signals. Samples from Gaussian processes have also been used to train deep learning methods, but again the samples used were perfectly periodic \cite{Agostinelli2016}. We tested the new method against the LSP, and while the the Gaussian process method produced a better FDR, the LSP has a faster computational runtime. The LSP may therefore be useful for preliminary analysis of a large dataset to get rapid results.

An alternative but related method was proposed \cite{Westermark2009}, which attempted to discriminate whether single cell circadian imaging data is best described by self-sustained limit cycle oscillations or by noise-induced oscillations. Whilst both methods assume that the dynamics are noisy, the data fitting procedure used by Westermark et al. (2009) is markedly different, as their model is fitted directly to the empirical autocorrelation of the data. This allows one to describe how well each model fits the data in isolation, but cannot make comparisons between models to ask which is more likely given the data. Here we employed Gaussian processes to find the probability of the data under two competing models (OU and OUosc), and this allowed us to make direct comparisons between the two using a statistical test on the LLR between the two models. 

As a practical application of their algorithm, Westermark et al. (2009) fitted circadian time series of cells with gene knock-outs and subsequently compare the parameters of the fitted models. If a genetic perturbation causes a systematic change to the dynamics then mutants may carry signatures in their time series that can be discovered by clustering in parameter space. This clustering of time series with parameters can also be performed with our method, as the period and quality parameters are estimated for each cell. Additionally, our method can also be used to quantify the percentage of cells with oscillatory versus non-oscillatory activity for various mutants. This allows quantitative assessment of the degree of disruption to oscillatory dynamics caused by a genetic mutation, which has a wide range of applications for both circadian and ultradian fields.

Our analysis method could form a starting point from which extensions can be designed to customise it for different experimental purposes. Firstly, our method could be extended to form a statistical test for whether two oscillating time series are coupled or synchronised. For example, dual-colour imaging in a single cell can be used to image the expression of two genes simultaneously. Using the same Gaussian process framework, one model could propose that the oscillations are independent while another could describe them as coupled oscillators (with a possible phase-shift between them). Through calculating an LLR between the competing models, a confidence measure could similarly be defined that the two oscillators are coupled.

Secondly, our method currently assumes that gene expression is either oscillatory or non-oscillatory for the entire duration of the time series. During development, however, stem cells may make dynamic transitions from oscillatory to non-oscillatory gene expression as they differentiate into more specialised cells \cite{Imayoshi2013, Goodfellow2014}. A ``changepoint'' model can be proposed in order to account for this behaviour, where the covariance function can switch between OU and OUosc at a given time point. Gaussian processes have previously been used for changepoint detection (e.g. \cite{Saatci2010}).

Finally, the method could be adapted to infer spatial organisation and pattern formation. The starting point of the method was a chemical reaction system in a well-mixed compartment, where space was neglected. In spatial systems, the combination of reaction and diffusion can lead to the formation of Turing Patterns \cite{Turing1952}, such as digit patterning during limb development \cite{Badugu2012}. The LNA can be applied to derive equations for stochastic Turing patterns \cite{Biancalani2010}, where intrinsic noise can induce pattern formation for parameters where the deterministic system reaches a homogeneous state. A similar method to the one presented here could therefore be applied to spatial systems to provide a confidence measure that a system is spatially organised in patterns as opposed to randomly structured \cite{Solin2013}. 

By quantifying the proportion of oscillating cells in a statistically objective manner, we envision that our method will provide a useful tool to single-cell gene expression community and it will be expanded upon for future applications.

\section*{Acknowledgments}

We are grateful to Prof. Mike White and Dr. David Spiller at the Systems Microscopy Suite for their invaluable help with imaging.

\newpage

\section*{Supporting information}

\beginsupplement

\newpage

\newpage

\newpage 

\newpage

\newpage

\newpage 

\newpage 

\newpage 

\newpage 

\newpage 

\newpage

\clearpage
\begin{table}[htbp]
\centering
\resizebox*{\textwidth}{\dimexpr\textheight-2\baselineskip\relax}{%
\begin{tabular}{|c|c|c|c|c|c|}
\hline
LLR (-4.5) & Cell number & LLR (-4) & Cell number & LLR (-5) & Cell number \\
\hline
 40.9984 & {\color{red} cell 13} &     40.9984 & {\color{red} cell 13} &   40.9984 & {\color{red} cell 13} \\
\hline
 33.8374 & {\color{red} cell 10} &      35.728 & {\color{red} cell 15} &   33.8374 & {\color{red} cell 10} \\
\hline
 31.0297 & {\color{red} cell 15} &     33.8374 & {\color{red} cell 10} &    30.075 & {\color{red} cell 15} \\
\hline
 28.2382 & {\color{red} cell 9 } &     28.8639 & {\color{red} cell 9 } &   25.4019 & {\color{red} cell 9 } \\
\hline
 20.9672 & {\color{red} cell 11} &     22.0586 & {\color{red} cell 11} &   16.6195 & {\color{red} cell 11} \\
\hline
 16.4446 & {\color{red} cell 12} &     18.5871 & {\color{red} cell 14} &   16.4446 & {\color{red} cell 12} \\
\hline
 14.9534 & {\color{red} cell 18} &     16.4446 & {\color{red} cell 12} &   13.4609 & {\color{red} cell 18} \\
\hline
 14.4372 & {\color{red} cell 14} &     14.8987 & {\color{red} cell 18} &   10.8794 & {\color{red} cell 17} \\
\hline
 11.8458 & {\color{red} cell 2 } &     12.9729 & {\color{red} cell 17} &   10.6184 & {\color{blue} cell 44} \\
\hline
 10.8794 & {\color{red} cell 17} &     12.3385 & {\color{red} cell 2 } &   10.2116 & {\color{red} cell 2 } \\
\hline
 10.6184 & {\color{blue} cell 44} &      10.727 & {\color{blue} cell 27} &   10.1728 & {\color{red} cell 14} \\
\hline
 9.14999 & {\color{blue} cell 27} &     10.6184 & {\color{blue} cell 44} &   7.75873 & {\color{blue} cell 26} \\
\hline
 8.75373 & {\color{red} cell 8 } &     8.88324 & {\color{red} cell 8 } &   7.71507 & {\color{red} cell 8 } \\
\hline
 7.87346 & {\color{blue} cell 26} &     8.00678 & {\color{blue} cell 26} &   5.76766 & {\color{blue} cell 28} \\
\hline
 7.21282 & {\color{red} cell 19} &     7.63687 & {\color{red} cell 19} &   5.33541 & {\color{red} cell 6 } \\
\hline
 5.64054 & {\color{red} cell 6 } &     7.36889 & {\color{blue} cell 42} &   4.56567 & {\color{red} cell 19} \\
\hline
 5.56432 & {\color{blue} cell 28} &     7.33647 & {\color{red} cell 4 } &   4.37002 & {\color{red} cell 3 } \\
\hline
  5.5251 & {\color{red} cell 4 } &     7.30956 & {\color{red} cell 5 } &   4.09345 & {\color{blue} cell 27} \\
\hline
 5.45709 & {\color{blue} cell 42} &     6.10228 & {\color{red} cell 6 } &    4.0546 & {\color{blue} cell 35} \\
\hline
 4.89476 & {\color{red} cell 5 } &     6.09968 & {\color{blue} cell 22} &   3.91524 & {\color{red} cell 4 } \\
\hline
 4.63679 & {\color{red} cell 3 } &     4.56951 & {\color{blue} cell 39} &   3.63046 & {\color{blue} cell 42} \\
\hline
 4.24299 & {\color{blue} cell 22} &     4.55454 & {\color{red} cell 3 } &   3.22251 & {\color{blue} cell 22} \\
\hline
 4.08948 & {\color{blue} cell 35} &     4.09915 & {\color{blue} cell 35} &   2.76633 & {\color{blue} cell 32} \\
\hline
 3.25271 & {\color{blue} cell 39} &     3.85603 & {\color{blue} cell 28} &   2.65737 & {\color{red} cell 7 } \\
\hline
 2.77461 & {\color{red} cell 7 } &     3.44772 & {\color{blue} cell 36} &   2.58263 & {\color{blue} cell 24} \\
\hline
 2.72791 & {\color{blue} cell 32} &     3.05206 & {\color{red} cell 7 } &   2.31273 & {\color{blue} cell 43} \\
\hline
 2.68822 & {\color{blue} cell 36} &     2.73167 & {\color{blue} cell 25} &   2.29182 & {\color{blue} cell 38} \\
\hline
 2.60088 & {\color{blue} cell 24} &     2.72791 & {\color{blue} cell 32} &    2.2635 & {\color{blue} cell 39} \\
\hline
 2.39872 & {\color{blue} cell 25} &     2.43911 & {\color{blue} cell 24} &    2.0984 & {\color{blue} cell 36} \\
\hline
 2.36568 & {\color{blue} cell 43} &     2.36568 & {\color{blue} cell 43} &   1.81549 & {\color{red} cell 16} \\
\hline
 2.22199 & {\color{blue} cell 38} &     2.31695 & {\color{blue} cell 38} &   1.72817 & {\color{red} cell 5 } \\
\hline
 1.87675 & {\color{red} cell 16} &     1.90276 & {\color{red} cell 1 } &   1.67971 & {\color{blue} cell 25} \\
\hline
 1.19864 & {\color{blue} cell 30} &     1.68928 & {\color{red} cell 16} &  0.880202 & {\color{blue} cell 30} \\
\hline
0.811085 & {\color{blue} cell 37} &     1.65826 & {\color{blue} cell 20} &  0.859294 & {\color{blue} cell 20} \\
\hline
 0.80946 & {\color{blue} cell 20} &     1.40088 & {\color{blue} cell 34} &  0.476036 & {\color{blue} cell 33} \\
\hline
0.708221 & {\color{red} cell 1 } &     1.37253 & {\color{blue} cell 30} &  0.310482 & {\color{blue} cell 34} \\
\hline
0.624682 & {\color{blue} cell 33} &     1.21808 & {\color{blue} cell 37} &  0.255126 & {\color{blue} cell 37} \\
\hline
0.494293 & {\color{blue} cell 31} &    0.997047 & {\color{blue} cell 31} &  0.073898 & {\color{blue} cell 41} \\
\hline
0.310482 & {\color{blue} cell 34} &    0.953552 & {\color{blue} cell 41} & 0.0212264 & {\color{blue} cell 29} \\
\hline
0.264505 & {\color{blue} cell 41} &    0.663787 & {\color{blue} cell 33} &         0 & {\color{red} cell 1 } \\
\hline
0.157088 & {\color{blue} cell 29} &    0.181401 & {\color{blue} cell 29} &         0 & {\color{blue} cell 21} \\
\hline
       0 & {\color{blue} cell 21} & 1.99944e-05 & {\color{blue} cell 40} &         0 & {\color{blue} cell 23} \\
\hline
       0 & {\color{blue} cell 23} &           0 & {\color{blue} cell 21} &         0 & {\color{blue} cell 31} \\
\hline
       0 & {\color{blue} cell 40} &           0 & {\color{blue} cell 23} &         0 & {\color{blue} cell 40} \\
\hline
\end{tabular}}
\caption{Ranked LLR list with different detrending parameters}
\label{table:MyTableLabel}
\end{table}
\clearpage


%
%
%

\end{document}